\documentclass{article}
\pdfoutput=1

\usepackage{geometry}
\usepackage{amsmath,amssymb}
\usepackage{multirow}
\usepackage{subcaption, array,booktabs}
\usepackage{hyperref}
\usepackage{adjustbox}
\usepackage{makecell}

\usepackage[utf8]{inputenc}
\DeclareUnicodeCharacter{2212}{-}
\usepackage{pgfplots}
\usepgfplotslibrary{groupplots,dateplot}
\usetikzlibrary{patterns,shapes.arrows}
\pgfplotsset{compat=newest}
\usepackage{stmaryrd, mathtools}
\usepackage{tikz}

\title{FBI: Fingerprinting models with Benign Inputs}
\author{Thibault Maho, Teddy Furon, Erwan Le Merrer\\
Centre Inria de l'Université de Rennes, France
}

\newcommand{\var}{$\mathcal{V_{P, \eta}}$}

\def \ie{\textit{i.e. }}
\def \eq{\textit{e.g. }}
\def \etc{\textit{etc}}

\begin{document}
\maketitle

\begin{abstract}

Recent advances in the fingerprinting of deep neural networks detect
instances of models, placed in a black-box interaction
scheme. Inputs used by the fingerprinting protocols are specifically
crafted for each precise model to be checked for. While efficient in
such a scenario, this nevertheless results in a lack of guarantee
after a mere modification (\eq retraining, quantization) of a model.

This paper tackles the challenges to propose i) fingerprinting schemes
that are resilient to significant modifications of the models, by generalizing to the notion of model
families and their variants, ii) an extension of the fingerprinting
task encompassing scenarios where one wants to fingerprint not
only a precise model (previously referred to as a \textit{detection}
 task) but also to identify which model family is in the black-box
(\textit{identification} task).

We achieve both goals by demonstrating that benign inputs, that are
unmodified images, for instance, are sufficient material for both
tasks. We leverage an information-theoretic scheme for the
identification task. We devise a greedy discrimination algorithm for
the detection task. Both approaches are experimentally validated over
an unprecedented set of more than 1,000 networks.
\end{abstract}
%

\def \det{\mathsf{D}}
\def \id{\mathsf{I}}
\newcommand{\set}[1]{\mathcal{#1}}
\def \setA{\set{A}}
\def \setX{\set{X}}
\def \setZ{\set{Z}}
\def \setC{\set{C}}
\def \setB{\set{B}}
\def \family{\set{F}}
\def \defi{\coloneqq}

\def \q{\mathbf{q}}
\def \ybv{z_{1:\ell}}
\def \mod{\mathsf{m}}
\def \var{\mathsf{v}}
\def \bb{\mathsf{b}}
\def \Vf{\mathsf{V}}
\def\acc{\mathsf{acc}}
\def \Exp{\mathbb{E}}

\newcommand{\yb}[1]{z_{{#1}}}
\newcommand{\ent}[1]{\llbracket #1\rrbracket}
\newcommand{\tuple}[2]{#1_{1:#2}}

\def \Prob{\mathbb{P}}
\def \sur{\mathsf{S}}
\def \tz{\tilde{z}}
\def \ty{\tilde{y}}
\def \tZ{\tilde{Z}}
\def \tY{\tilde{Y}}
\section{Introduction}
Fingerprinting classifiers aim at deriving a signature uniquely identifying a model, like the human fingerprint's minutiae in biometry.
This is essentially a black-box problem: the classifier to be identified is in a black-box in the sense that one can just make some queries and observe its outputs.
For instance, the model is embedded in a chip or we can query it through an API (MLaaS).

The main application that related works \cite{9401119,ipguard,arxiv.2202.08602,pan2021tafa,ZHAO2020488} target is the proof of
ownership. An accurate deep neural network is a valuable industrial
asset due to the know-how for training it, the difficulty of gathering
a well-annotated training dataset, and the required computational
resources to learn its parameters. The cost of GPT-3, one of the
biggest and most accurate NLP models is estimated to 4.6 million
dollars\footnote{https://lambdalabs.com/blog/demystifying-gpt-3/}.
In this context, the entity identifying a black-box wants to \emph{detect}
whether it is not a stolen model of her.

Another at least as critical application is information gain.
For instance, an attacker willing to delude the
classifier first gains some knowledge about the remote model, or a
company wants to determine which model is in use in a competitor's
production system. This application has been left aside as of today,
and we tackle it under the notion of the fingerprinting \emph{identification} task.

For clarity, we name Alice the entity willing to identify the model that Bob has embedded in the black-box.

\paragraph{Challenges} The biggest difficulty is that there exist plenty of ways to modify a model while maintaining its intrinsic good accuracy.
These procedures simplify a network (quantification of the weights
and/or activations, pruning, see \eq \cite{han2015deep}), or make it
more robust (preprocessing of the input, adversarial re-training
\cite{ganin2016domain}).
We hereafter name a modified model a \emph{variant}.
These mechanisms were not a priori designed to make fingerprinting harder but they leave room for Bob to tamper with the fingerprint of a model.
We assume that Alice also knows some of these procedures.
Yet, they are often defined by many parameters and among them scalars so that  there is virtually an infinite continuum of variants.
Like in biometry, the fingerprint should be discriminative enough for being unique per model but also sufficiently robust to identify a variant. 

\begin{figure}[bt]
\begin{center}
		\resizebox{0.9\linewidth}{!}{\input{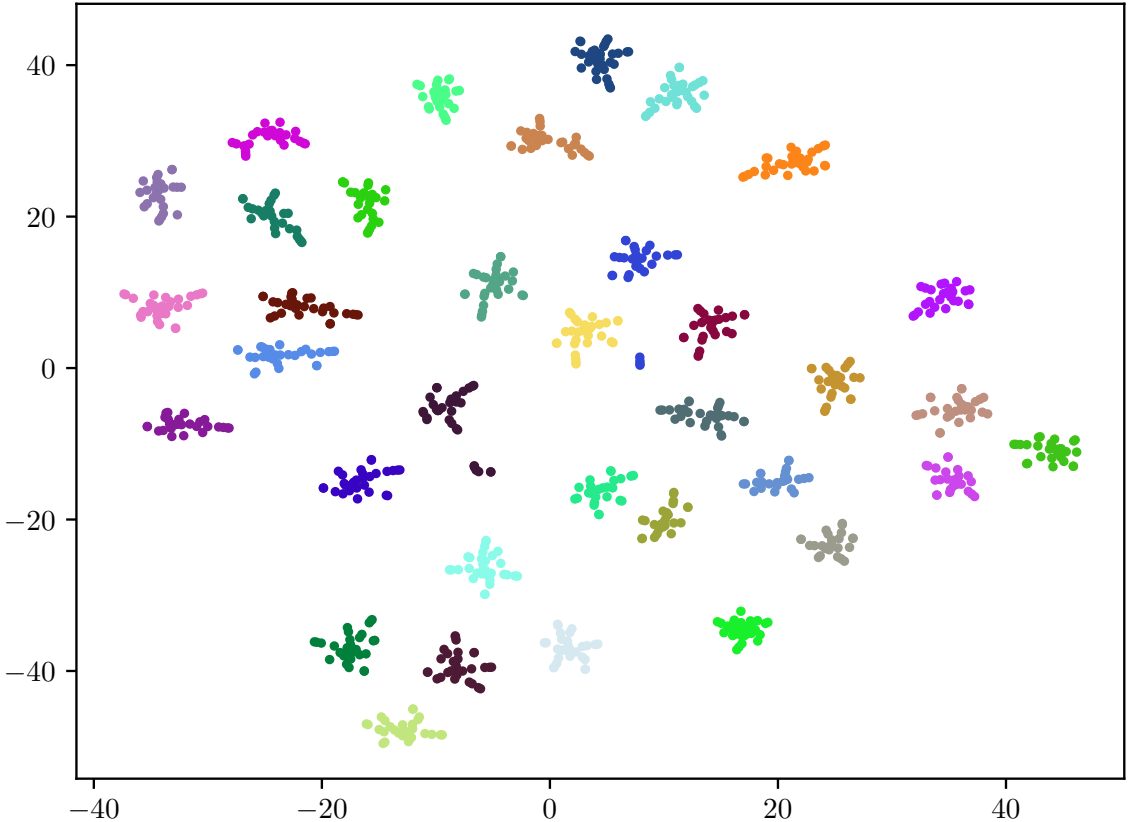}}
\caption{A t-SNE representation of the pairwise distances of 1081 different models: 10 types of variation applied on 35 off-the-shelves vanilla models for ImageNet with different parameters (listed in App.~\ref{app:DistanceModel}).
This work exploits the clear separability (clusters of consistent colors) observed in the decisions of these models. Confusions yet happen (model colors further apart from their cluster), but are under scrutiny for the tracking of false positive identification.}
\label{fig:Appetizer}
\end{center}
\end{figure}

The approaches in the literature have two noticeable common pillars.
They use the frontiers in the input space drawn by a classifier as the fingerprint, \ie the unique signature identifying the model \cite{ipguard, pan2021tafa, ZHAO2020488}.
Two neural networks sharing the same architecture, the same training set, and the same training procedure are different because the training is stochastic (like the Stochastic Gradient Descent). This makes their frontiers in the input space not overlap.
Most of the papers in the literature are looking for discriminative deviations of these frontiers.
Second, the key task is detection: Alice makes a guess about the model in the black-box and then she queries these specific queries to test whether her hypothesis holds~\cite{ipguard,sensitive,pan2021tafa,ZHAO2020488}.

\paragraph{Rationale}
We find that before resorting to the crafting of inputs for
fingerprinting models, the use of benign inputs
has not been thoroughly investigated. The use of benign inputs would
constitute a certain advantage, as it would remove the need for the
design of often complex ways to efficiently craft these.
It is less prone to defenses being implemented on Bob's side (\eq rejection
based on the distance to the decision frontier \cite{meng2017magnet}).  A
second salient observation is the restriction of previous works on the
detection task. The more general possibility to identify a model or a
family of models inside the black-box remains unstudied. Our work thus
differs from related works on these two key aspects: \textit{i)} we do
not forge any specific input but use regular benign inputs (so that we
do not need to probe the input space to discover the decision
frontiers), and \textit{ii)} we directly identify models using their intrinsic classification
difficulties on a handful of inputs, this is more efficient than sequentially detecting models until finding the good one.

In a nutshell, when Bob has picked a model among a set of networks
known by Alice, then our solution is essentially deterministic: Alice
has to find the sequence of inputs of minimum length to identify the
black-box. We apply a greedy algorithm that carefully selects the
input to iteratively narrow down the set of suspects, \ie candidate
models, until it becomes a singleton. Approximation theory tells that
this is suboptimal but we report that in practice many networks are
indeed identified within less than three queries.

When an attack by Bob has made a variant of a model, the output of it
may not match the output of any known model. We then use
C.E. Shannon's information theory to measure the statistical
similarity between the outputs of two models.
As an appetizer, Figure~\ref{fig:Appetizer} depicts the
t-SNE representation from the pairwise distances within a
set of 35 vanilla models and their variants.
The model families are well clustered in the sense that
variants are closer to their original network than any other model.
Alice may not identify precisely the variant of the model but at least
she can identify its family, \ie infer which was the original vanilla
model and even which kind of variation was applied to it.

\paragraph{Contributions} Our contribution is fourfold.\\
1) We demonstrate that the mere use of benign images is enough to
  accomplish high success rates for fingerprinting modern models.
  This is to be opposed to the computationally demanding task of
  crafting inputs for that same goal.\\
2) The fingerprinting detection task, introduced by
  state-of-the-art works, is complemented with the introduction of the
  identification task. We frame the latter as an information
  theoretical problem.\\
3) We present a distance based on the empirical Mutual Information,  gauging how close two models are.
This distance permits to generalize the notion of attacks on models through the notion of model families and variants.\\
4) We perform extensive experimentation by considering more than
  1,000 classification models on ImageNet. A head-to-head comparison
  with the two related works reports significant improvements
  w.r.t. accuracy in the detection task.

\paragraph{Document structure}
Section~\ref{sec:ThreatModel} is a threat analysis listing all the
working assumptions in our work.  The next two sections deal both with
detection (Alice verifies her hypothesis about the black-box) and
identification (Alice discovers which model is in the black-box) but
under two scenarios: Section~\ref{sec:known} builds on the fact that
Bob has picked a model among the set known by Alice, whereas
Section~\ref{sec:unknown} assumes that the black-box may be an unknown model.
Both sections contain experimental
results. Section~\ref{sec:SOTA} is devoted to the related works and
the benchmark with state-of-the-art detection schemes.
The appendices contain a summary of the notations, details about the experimental protocol, and supplementary results.


\section{Threat Model}
\label{sec:ThreatModel}
This section details the goals of Alice and Bob.

\subsection{Bob: keeping  his model anonymous}

\subsubsection{Goals}
Bob is playing first by secretly selecting a model and putting it in
the black-box. This model can be a vanilla model or a variant of a
known model.  A variant is created by applying on a given vanilla
model $\mod$ the procedure $\Vf$ parametrized by $\theta\in\Theta$
which describes the type of modification and the associated
parameters; this can be thought of as an attack by Bob on the vanilla
model to harden identification.  We denote such a variant by $\var
= \Vf(\mod,\theta)$.

The goal of Bob is to offer an accurate black-box classifier while
maintaining the `anonymity' of the model in use.  The first
requirement is that a small loss in performance is tolerated. If a
variant does not comply with this criterion then Bob cannot consider
it as an option.  In classification, the performance of a model $\mod$
is often gauged by the top-1 accuracy, denoted $\acc(\mod)$.  We
formalize this requirement as
\begin{equation}
\label{eq:criterion}
\frac{\acc(\mod) - \acc(\Vf(\mod,\theta))}{\acc(\mod) } <\eta,
\end{equation}
 where $\eta>0$ is the tolerance (set to 15\% in our experimental work).

\subsubsection{Ressources}
The second requirement is more delicate. We first need to limit the power of Bob.
If Bob creates an accurate model \textit{ex nihilo}, then Alice can pursue neither detection nor identification.
We assume that Bob cannot train such a model from scratch because he lacks good training data, expertise in machine learning, or computing resources.  
This also means that Bob cannot retrain a model, or only up to a limited extent.
In other words, the complexity of the procedure creating $\var = \Vf(\mod,\theta)$ ought to be much smaller than the effort spent at training the original model $\mod$.

Our experimental work considers two kinds of procedures:\\
1) modification of the input: $\var(x) = \mod(\mathsf{T}(x,\theta))$. Classifiers are robust to benign transformations of the input. As far as images are concerned, the transformation $\mathsf{T}$ can be JPEG compression, histogram equalization, blurring, \etc. In the same spirit, Randomized Smoothing consists in adding noise patterns to the input and aggregating the predicted classes into one single output.\\
2) modification of the model: $\var(x) = \mathsf{T}(\mod,\theta)(x)$: The transform $\mathsf{T}$ slightly changes the model weights by quantization, pruning, adversarial retraining, finetuning... Some of these procedures require small retraining with few resources so as not to lose too much accuracy.

In the sequel, the model in the black-box is denoted by $\bb$ and $\setB$ is the set of possibilities: $\bb\in\setB$.
This set is defined as:
\begin{equation}
\label{eq:setB}
\setB\defi\left\{\var = \Vf(\mod,\theta): \mod\in\set{P}, \theta\in\Theta, \acc(\var)>(1-\eta)\acc(\mod) \right\},
\end{equation}
where $\set{P}$ is a set of vanilla models and $\Theta$ a set of transformations (encompassing the identity $\var=\mod$).

\subsection{Alice: disclosing the remote model}

\subsubsection{Goals}
The task of Alice is to disclose which model is in the black-box. This has two flavours: detection or identification.

\textit{Detection} (denoted by $\det$) means that Alice performs a hypothesis test. She
first makes a hypothesis about the black-box, then makes some
queries, and finally decides whether the hypothesis holds based on the
outputs of the black-box.  The outcome of the detection is thus
binary: Alice's hypothesis is deemed correct or not. This is the
nominal use case in the related works \cite{9401119,ipguard,arxiv.2202.08602,pan2021tafa,ZHAO2020488}.

\textit{Identification} (denoted by $\id$) means that Alice has no prior about the model
in the black-box.  She makes queries and processes the outputs to
finally make a guess.  The outcome is either the name of a model she
knows, or the absence of a decision if she has not enough evidence.

\subsubsection{Knowledge about the black-box}
\label{sec:knowledge_alice}
The second crucial point is her knowledge about the black-box.  Alice
can only detect or identify a relation to a model she knows: this
means that she has an implementation of this model, which she can
freely test in white-box access, typically locally on her machine.
We denote the set of models known by Alice by
$\setA$.

As by the very definition of a \emph{variant}, Alice may know some of
them but not all of them. For instance, some procedures $\Vf$ admit a
real number as a parameter.  Therefore, there is virtually an infinite
number of variants. This leads to the convenient notion of a model
\emph{family}, we now introduce under three flavours:
\begin{itemize}
\item $\family(\mod)$: This family is the set of all variants made from the original vanilla model $\mod$:
\begin{equation}
\family(\mod)\defi\left\{v=\Vf(\mod,\theta): \theta\in\Theta\right\}.
\label{eq:Family1}
\end{equation}
\item $\family(\mod,\Psi)$: This family is the set of all variants made from the original vanilla model $\mod$ by a specific procedure:
\begin{equation}
\family(\mod,\Psi)\defi\left\{v=\Vf(\mod,\theta): \theta\in\Psi\subset\Theta\right\},
\label{eq:Family2}
\end{equation}
where $\Psi$ denotes the subset of parameters related to this specific procedure.
\item $\family(\mod, \{\theta\})$: This family is a singleton composed of a particular variant:
\begin{equation}
	\family(\mod, \{\theta\})\defi\left\{v=\Vf(\mod,\theta)\right\}.
	\label{eq:Family3}
\end{equation}
\end{itemize}
With these definitions in mind, detection is based on the hypothesis that the black-box belongs to a given family, while
Identification looks for the family the black-box belongs to.

\subsubsection{Resources}
A third element is the resources of Alice.
We already mention the set $\setA$ containing some vanilla models and few variants of theirs.
She also has a collection of typical inputs, \ie a testing dataset.
We suppose that they are statistically independent and distributed as the data in the training set of the models.
In the sequel, the collection of inputs is denoted $\setX = \{x_1,\ldots,x_N\}$.

In the end, be it for detection or identification, Alice selects some elements of $\setX$ for querying the black-box.
We denote this by an ordered list of indices: $q_{1:\ell} = (q_1,\ldots,q_\ell)\in\ent{N}^\ell$, where $\ent{N}\defi\{1,\ldots,N\}$.
This means that Alice first queries $x_{q_1}$, and then $x_{q_2}$ and so forth.
The outputs of the black-box are denoted as $\ybv = (\yb{q_1},\ldots,\yb{q_\ell})$, with $\yb{q_i} = \bb(x_{q_i})$.

\subsection{The classifier in the black-box}
The black-box works as any classifier.
We denote the set of possible classes $\setC$.
The output $\yb{}=\bb(x)$ for input $x$ is the first $k$ classes ordered by their predicted probabilities (\ie the top-$k$).
It means that $\yb{}$ is an ordered list in $\setC^k$: $\yb{}=(c_1,\ldots,c_k)$.
The set $\setZ_k$ of possible outcomes has a size as big as $(C)_k \defi C(C-1)\ldots(C-k+1)$.
The black-box only discloses these classes, not the associated predicted probabilities.
In the experimental work, the size of $\setC$ is 1,000 (ImageNet) and $k\in\{1,3,5\}$ which is usual in several image classification APIs.
We assume that the considered models and variants have an accuracy which is not perfect.
The top-1 accuracy of ImageNet classic models ranges from 70\% to 85\%.

\subsection{Summary}
This paper considers scenarios which are labeled as $(\mathsf{Task},\family,\setA,k)$ where $\mathsf{Task}\in\{\det,\id\}$ (Detection or Identification), $\family$ is the kind of family that will be inferred by Alice, $\setA$ is the set of models known by Alice, and $k$ indicates that the output of the black-box is the top-$k$ classes. 
There is a clear cut between the following two cases:\\
1) Walled garden: $\setA=\setB$. We impose that the black-box is one of the networks known by Alice.\\
2) Open world: $\setA\subsetneq\setB$. The black-box may not be a model known by Alice. This is the case when Bob uses an unknown variant, for instance.

This distinction drives the structure of the next sections because our solutions are of different nature.

\section{The black-box is a known model}
\label{sec:known}

Under the assumption that $\setA=\setB$, Alice achieves her goal when she correctly guesses which family the black-box belongs to.
The alternative is to fail to gather enough evidence to make a decision.
For the sake of clarity, we explain our procedure for a given family $\family\subset\setA$, which can be one of the three types of families presented in Sect.~\ref{sec:knowledge_alice}.

Alice has a set of models composed of some vanilla models $\set{P} = \{\mod_1,\ldots,\mod_M\}$ and some variants of theirs.
Alice also has the collection of benign inputs $\setX=(x_1,\ldots,x_N)$.
Offline, she creates a database of $|\setA|N$ outputs $ (\mod(x_j))_{\mod\in\setA,j\in\ent{N}}$.

Let $\set{D}$ be a subset of $\setA$. We define by $\set{D}(x) \defi \{\mod(x) | \mod \in \set{D}\}$ the set of labels predicted by the models in $\set{D}$ for input $x$.
With abuse of notations, $\set{D}(x_{\tuple{q}{\ell}})$ is the set of the concatenation of labels predicted by the models in $\set{D}$ for the entries $(x_{q_1},\ldots,x_{q_\ell})$.
Conversely, $\set{M}(x,y,\set{D})\defi \{\mod | \mod \in \set{D}, \mod(x) = y\}$ lists the models in $\set{D}$ predicting $y$ when $x$ is submitted.

\subsection{Detection $(\det,\family,\setA=\setB,k)$}
Alice first makes a hypothesis about a family $\family$, and her goal is to discover whether the outcome is \emph{positive} ($\bb \in \family$)  or \emph{negative} ($\bb \in \setA\backslash\family$).
We assume that Alice is convinced about her hypothesis and that she hopes for a \textit{positive}.
Our procedure thus focuses on reducing the number of models in $\set{A}\backslash\family$ likely to be the black-box.

Alice uses a greedy algorithm which leverages the information about the black-box retrieved from the previous queries.
According to the outputs of the black-box, several models can be discarded.
At step $\ell$,  $(\setA \backslash\family)^{(\ell)}$ (resp. $\family^{(\ell)}$) denote the subset of models in $\setA \backslash\family$ (resp. $\family^{(\ell)}$) which agree with the previous outputs. These are candidates in the sense that they could be the black-box model.
\begin{equation}
\label{eq:Greedy1}
(\setA \backslash\family)^{(\ell)} \defi \bigcap_{i=1}^\ell \set{M}(x_{q_i},\bb(x_{q_i}),\set{\setA \backslash\family}).
\end{equation}
Initially, all the models are candidates: $(\setA \backslash\family)^{(0)} = \setA \backslash\family$ and $\family^{(0)}=\family$.
At step $\ell+1$, the greedy algorithm sorts the inputs that have not yet been queried according to a score.
This score $s^{(\ell+1)}(x)$ reflects how much the set of candidates $(\setA \backslash\family)^{(\ell)}$ reduces if input $x$ is submitted next. 
It is thus based on the quantities $|\set{M}(x,y,(\setA \backslash\family)^{(\ell)})|$.

\paragraph{Expectation}
Convinced that her hypothesis is correct, Alice computes the expectation of the number of candidate models outside the family after querying input $x$.
The average is weighted by the number of models in $\family^{(\ell)}$ predicting a particular label, as if the black-box were randomly picked in $\family^{(\ell)}$:
\begin{equation}
s^{(\ell+1)}(x) = \sum_{y \in \family^{(\ell)}(x)} \left|\set{M}(x, y, (\setA \backslash\family)^{(\ell)})\right| \frac{|\set{M}(x, y, \family^{(\ell)})|}{|\family^{(\ell)}|}
\label{eq:AverageScoreDet}
\end{equation}
\paragraph{Worst Case}
Instead of a sum, Alice takes the maximum which corresponds to the worst case, at the same time likely to happen and reducing the list of candidates by a small amount if any:
\begin{equation}
s^{(\ell+1)}(x) = \max_{y \in \family^{(\ell)}(x)} \left|\set{M}(x, y, (\setA \backslash\family)^{(\ell)})\right| \frac{|\set{M}(x, y, \family^{(\ell)})|}{|\family^{(\ell)}|}.
\label{eq:MaxScoreDet}
\end{equation}

Alice then submits one of the inputs with the lowest score:
\begin{equation}
	q_{\ell+1} \in \arg\min_{k} s^{(\ell+1)}(x_k).
\end{equation}
Our procedure stops after $L$ iterations when meeting one of the three stopping criteria:
\begin{itemize}
\item $(\setA \backslash\family)^{(L)}=\emptyset$: The detection result is \emph{positive}. No model outside the family responds like the black-box.
The black-box is part of the family since we assume it belongs to $\setA$.
\item $\family^{(L)}  = \emptyset$: The detection result is \emph{negative}. The responses of the black-box are different from the ones of the models in the family $\family$.
\item $\min_{k} s^{(L+1)}(x_k) = |(\setA \backslash\family)^{(L)}|$: The detection failed. All the remaining models in $(\setA \backslash\family)^{(L)}$ and in $\family^{(L)}$ produce the same prediction no matter which input is submitted. It is therefore impossible to discern them.
\end{itemize}

\subsection{Identification $(\id,\family,\setA=\setB,k)$}
Identification means that Alice makes a partition of her set of models in disjoint families: $\setA = \cup_{i=1}^{n_\family} \family_i$.
Her goal is to identify which family the black-box belongs to.

It is easy to base a verification procedure onto a detection scheme assuming there is no failure.
Alice arbitrarily orders the families, and sequentially tests the hypotheses until she finds a match.
The expected number of queries is given by (see the proof in App.~\ref{sec:ProofEqIdeExp}):
\def \negg{\mathsf{neg}}
\def \poss{\mathsf{pos}}

\begin{equation}
\label{eq:IdeExp}
\Exp(L) = \frac{1}{n_{\family}}\sum_{j=1}^{n_\family}\Exp\left(L^\poss_j\right)  + \frac{n_\family-1}{2n_\family} \sum_{j=1}^{n_\family}\Exp\left(L^\negg_j\right)
\end{equation}
where $(\Exp(L^\poss_j),\Exp(L^\negg_j))$ are the expected number of queries necessary for taking a positive or negative decision about the \emph{detection} of the hypothesis $\family_j$.

We propose a better approach based on a greedy algorithm similar to the detection one.
Suppose that Alice has already submitted $\ell$ queries to the black-box.
By comparing the outputs of the black-box and of the models she knows, she is able to distinguish models which are not in the black-box from models likely to be in the black-box. This list of remaining models is denoted $\setA^{(\ell)}$.
The goal of Alice is to reduce the set of candidates to a single family $\family_i$, not knowing in advance the model Bob placed in the black-box:
\begin{equation}
	\exists i, \family_i \subset \setA^{(\ell)}.
\end{equation}

In the beginning, all the models are possibly in the black-box, \ie\ $\setA^{(0)} = \setA$.
At step $\ell+1$, the greedy algorithm chooses the best input to query next knowing $\setA^{(\ell)}$.
Alice may resort to the following heuristics.

\paragraph{Expectation.} Alice supposes that the black-box is randomly chosen uniformly in the set of remaining models $\setA^{(\ell)}$.
For any input $x$ not queried yet, she computes the expectation of the numbers of remaining families if that input were selected next, \ie $\{\family_i | \family_i \cap \setA^{(\ell + 1)} \neq \emptyset \}$.
She randomly chooses among the inputs which minimize this figure:
\begin{equation}
	s^{(\ell+1)}(x) = \sum_{y \in \setA^{(\ell)}(x)} \left[ \sum_{i=1}^{n_\family} \delta\left(\set{M}(x, y, \family_i^{(\ell)}) \neq\emptyset\right) \right] \frac{|\set{M}(x, y, \setA^{(\ell)})|}{|\setA^{(\ell)}|},
	\label{eq:AverageScoreIde} 
\end{equation}
where $\delta$ is the indicator function of an event.

\paragraph{Worst case.} For any input not queried yet, Alice computes the maximum of the numbers of remaining families if that input were selected next weighted by the probability of this event:
\begin{equation}
	s^{(\ell+1)}(x) = \max_{y \in \setA^{(\ell)}(x)} \left[ \sum_{i=1}^{n_\family} \delta\left(\set{M}(x, y, \family_i^{(\ell)}) \neq\emptyset\right) \right] \frac{|\set{M}(x, y, \setA^{(\ell)})|}{|\setA^{(\ell)}|}
	\label{eq:MaxScoreIde} 
\end{equation}

The input to be submitted is randomly picked  among the ones with the lowest score:
\begin{equation}
	q_{\ell+1} \in \arg\min_{k} s^{(\ell+1)}(x_k).
\end{equation}

\subsection{Experimental work}
\label{sec:ExpWork1}
\subsubsection{Detection}
A first experimental work  measures the number of queries needed for detection with the three types of family defined in Eq.~\eqref{eq:Family1}, \eqref{eq:Family2} and \eqref{eq:Family3}.
It considers two cases: Alice's hypothesis is correct (positive case) or incorrect (negative case).
The cases are not analyzed exhaustively.
For example, in the negative case for a singleton family (\ie\ Alice is wrong to suspect that the black-box is $\family(m,\{\theta\})$),
there are $|\setA|(|\setA|-1)$ possible combinations, \ie more than a million. 
Instead, the experiment randomly picks 1.000 positive and 1.000 negative among all these cases. 

Figure~\ref{fig:histogram_all_known_detection} shows the best results obtained with the expectation score~\eqref{eq:AverageScoreDet}.
Results for the worst-case score~\eqref{eq:MaxScoreDet} are shown in App.~\ref{app:known}, Fig.~\ref{fig:histogram_all_known_detection_worst_case}.
As a side-product, Table~\ref{tab:detection_single_image_count} gives the percentage of inputs in $\setX$ answering the detection problem under the best case, \ie when one unique query is sufficient. 

\begin{figure*}
	\centering
	
	\resizebox{0.32\linewidth}{!}{\input{./images/omniscient/detection/score_mean-case-pure.pgf}}
	\resizebox{0.32\linewidth}{!}{\input{./images/omniscient/detection/score_mean-case-variation.pgf}}
	\resizebox{0.32\linewidth}{!}{\input{./images/omniscient/detection/score_mean-case-singleton.pgf}}
	
	\caption{\label{fig:histogram_all_known_detection}Probability distribution of the number of queries for $(\det,\family,\setA=\setB,k)$ using the expectation score~\eqref{eq:AverageScoreDet} when the black box returns top-$k$ classes with $k=1$  (\textit{blue}), $k=3$ (\textit{red}) or $k=5$ (\textit{green}). Family considered from left to right: $\family(\mod)$~\eqref{eq:Family1}, $\family(\mod,\Psi)$~\eqref{eq:Family2}, and $\family(\mod,\{\theta\})$~\eqref{eq:Family3}.}
\end{figure*}

\begin{table}[b]
	\centering
	\caption{\label{tab:detection_single_image_count} Percentage of inputs in $\setX$ concluding detection $(\det,\family,\setA=\setB,k)$ within a single query.}
	\begin{tabular}{l|ccc}
		\toprule
		Family & top-1 & top-3 & top-5\\
		\midrule
		Vanilla $\family(\mod)$~\eqref{eq:Family1}		& $0.28\%$ & $1.2\%$  & $\bf{12.5\%}$ \\
		Variation $\family(\mod,\Psi)$ ~\eqref{eq:Family2}  	& $0.31\%$ & $4.8\%$  & $\bf{21.0\%}$ \\
		Singleton $\family(\mod,\{\theta\})$~\eqref{eq:Family3} & $0.37\%$ & $8.3\%$  & $\bf{31.4\%}$ \\
		\bottomrule
	\end{tabular}
\end{table}

\paragraph{Few queries are enough.}
When the detection succeeds (with a positive or negative decision about the hypothesis), at most three queries are needed, and in most cases, only one is sufficient. This holds although the greedy algorithm is known to be suboptimal.
When the greedy algorithm needs three inputs, another algorithm may only need two queries.
Yet, when the greedy algorithm needs two, no algorithm can do better because the greedy would have found an unique input if existing.
Positive and negative conclusions are roughly drawn within the same number of queries, although our algorithm is designed to quickly prove positive detections.

\paragraph{Few failures.} The inability to detect the model in the black-box as part of the family $\family$ happens when:
\begin{equation}
\label{eq:ExplainFail}
	\exists \mod \in \family, \exists \mod' \in \setA \backslash \family, \forall x\in\setX,~~\mod(x) = \mod'(x).
\end{equation}
The failures occur when the family corresponds to a set of variations~\eqref{eq:Family2} or an exact model~\eqref{eq:Family3}.
It happens that the algorithm cannot distinguish a few pairs of different variations issued from the same vanilla model.
This is the only possible explanation: Otherwise, \ie the indistinguishable models come from two different vanilla networks, this setup would also end up in a failure when detecting families spanned from a vanilla model~\eqref{eq:Family3}, which is not reported in Fig.~\ref{fig:histogram_all_known_detection}.
In other words, Alice can always guess that the black-box is a variation of a given vanilla model, and rarely she cannot guess which variation it is exactly.

On the other hand, failures should also happen in the negative case. None is reported in Fig.~\ref{fig:histogram_all_known_detection} because they are statistically rare. For a given family $\family$, suppose that the models $\mod$ and $\mod'$ in~\eqref{eq:ExplainFail} are both unique.
A failure happens in the positive case if Bob puts model $\mod$ in the black-box. This happens with probability $1/|\family|$.
A failure happens in the negative case if Bob puts model $\mod'$ in the black-box. This happens with probability $1/|\setA \backslash \family| < 1/|\family|$. 

Experimentally, the number of queries to end up in a failure is similar to the number of queries for getting a positive outcome.

\paragraph{A bigger top-$k$ is better.}
When the output of the black-box is rich, \ie top-$k$ classes with $k>1$, one unique input is sufficient.
Moreover, Table~\ref{tab:detection_single_image_count} shows that there are more of these unique inputs in $\setX$.
In this case, Alice no longer needs a large collection of benign inputs. 

\paragraph{A bigger family is harder to detect.}
Families of type~\eqref{eq:Family1} are bigger than families~\eqref{eq:Family2} which are bigger than the singleton~\eqref{eq:Family3}.
Ignoring the failure case, Figure~\ref{fig:histogram_all_known_detection} and Table~\ref{tab:detection_single_image_count} show that it is harder to detect a large family. It is more frequent that some model members take different outputs in large families.
On the contrary, we observe that the variants of the same model with the same variation but with different parameters often share the same output.

\subsubsection{Identification}
The protocol is similar to the previous one for detection. 
Figure~\ref{fig:histogram_all_known_identification} shows the best results obtained with the expectation score~\eqref{eq:AverageScoreIde}.
The results for the other score~\eqref{eq:MaxScoreIde} are slightly worse and shown in App.~\ref{app:known}, Fig.~\ref{fig:histogram_all_known_identification_worst_case}.

\begin{figure*}
	\centering
	
	\resizebox{0.32\linewidth}{!}{\input{./images/omniscient/identification/pure-score_mean-case.pgf}}
	\resizebox{0.32\linewidth}{!}{\input{./images/omniscient/identification/variation-score_mean-case.pgf}}
	\resizebox{0.32\linewidth}{!}{\input{./images/omniscient/identification/singleton-score_mean-case.pgf}}
		
	\caption{\label{fig:histogram_all_known_identification} Probability distribution of the number of queries for $(\id,\family,\setA=\setB,k)$ using the expectation score~\eqref{eq:AverageScoreDet} when the black box returns top-$k$ classes with $k=1$  (\textit{blue}), $k=3$ (\textit{red}) or $k=5$ (\textit{green}). Family considered from left to right: $\family(\mod)$~\eqref{eq:Family1}, $\family(\mod,\Psi)$~\eqref{eq:Family2}, and $\family(\mod,\theta)$~\eqref{eq:Family3}.}
\end{figure*}

\begin{figure*}
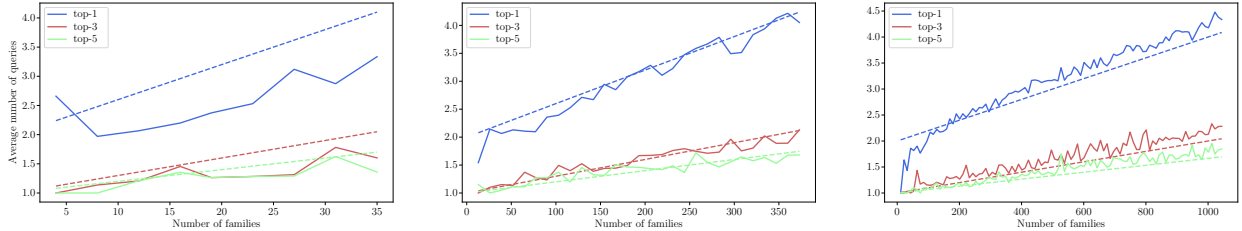

	\centering

	\resizebox{0.32\linewidth}{!}{\input{./images/omniscient/identification_mean_queries_n_families_with_regression/pure-score_mean-case.pgf}}
	\resizebox{0.32\linewidth}{!}{\input{./images/omniscient/identification_mean_queries_n_families_with_regression/variation-score_mean-case.pgf}}
	\resizebox{0.32\linewidth}{!}{\input{./images/omniscient/identification_mean_queries_n_families_with_regression/singleton-score_mean-case.pgf}}	
	
	\caption{\label{fig:all_known_identifcation_n_queries_n_families} Average number of queries as a function of the number of families $n_\family$ for $(\id,\family,\setA=\setB,k)$ with the expectation score~\eqref{eq:AverageScoreDet} and when the black-box returns top-$k$ classes with $k=1$  (\textit{blue}), $k=3$ (\textit{red}) or $k=5$ (\textit{green}). The dotted lines are the linear regressions. Family considered from left to right: $\family(\mod)$~\eqref{eq:Family1}, $\family(\mod,\Psi)$~\eqref{eq:Family2}, and $\family(\mod,\{\theta\})$~\eqref{eq:Family3}.}
\end{figure*}

\paragraph{Identification vs. Detection.} Comparing Figures~\ref{fig:histogram_all_known_detection}  and~\ref{fig:histogram_all_known_identification}, two times more queries are necessary for identifying a family rather than detecting it.
It is possible to identify a model quickly with at most five benign queries which are a lot less than the sequential procedure~\eqref{eq:IdeExp}.
Identification is a harder task than detection to a small extent.

The biggest difference is under the top-1 scenario where a unique query is rarely sufficient.
The 35 vanilla models considered here were trained on the same dataset. They have good accuracy ($>70\%$).
If many unique inputs to identify existed, this would mean that for any of these inputs, the 35 models give 35 different top-1 predictions.
Assuming that one of these models makes a correct classification, the other 34 models are wrong.
If a lot of these inputs existed, this would imply models with low accuracy.
In other words, these inputs are necessarily rare, or even non-existing.

\paragraph{A bigger top-$k$ is better.} In contrast to detection, the gain of information provided by top-3 and top-5 outputs is very substantial.
When the top-5 is returned, $90\%$ of the families are identified within one query.
The supervised training of the vanilla models only focuses on the top-1 s.t. it agrees with the ground truth class.
For $k>1$, the top-$k$ is almost specific of the network. This explains the big improvement from top-1 to top-$k$.

\paragraph{Number of families.} Figure~\ref{fig:all_known_identifcation_n_queries_n_families} represents the evolution of the average number of queries to identify one out of $n_\family$ families. The more families, the bigger the number of queries on average.
But this number also depends on the size of the families and the top-$k$.
We observe that the increase is roughly linear (see dashed lines in Fig.~\ref{fig:all_known_identifcation_n_queries_n_families}).
As a rule of thumb, we observe that the expectation of the number of queries roughly follows the empirical law: 
\begin{equation}
	\Exp(L) \approx 0.002 \times\frac{\Exp(|\family|) n_{\family}}{k} + \beta(k),
\end{equation}
where $\Exp(|\family|)$ is the average number of elements in the family.
This is a major improvement w.r.t.~\eqref{eq:IdeExp}.
For instance, for singleton family, $\Exp(|\family|) = 1$ and the rate equals $0.002$ under top-1, whereas the rate in~\eqref{eq:IdeExp} cannot be lower than $0.5$ since we need at least one query to discard a hypothesis, \ie $\Exp(L^\negg_j)\geq 1$.


\section{The black-box is an unknown model}
\label{sec:unknown}
This section assumes that $\setA\subsetneq\setB$ because $\setB$ contains unknown models or variants of some models known by Alice.

\subsection{Modeling}
\subsubsection{Assumptions}
\label{sec:unknown_assumptions}
Our working assumptions are the following: When queried by random inputs, a variant $\Vf(\mod,\theta)$ produces outputs statistically
\begin{itemize}
	\item independent from the outputs of a different model $\mod^\prime\neq\mod$.  
	\item dependent from the outputs of the original model $\mod$.  
\end{itemize}
We consider a particular procedure for generating a variant as being like a transmission channel.
The output $Z$ of the variant $\Vf(\mod,\theta)$ is as if the output $Y$ of the original model $\mod$ were transmitted to Alice through a noisy communication channel parametrized by $\theta$. Like in C.E. Shannon's information theory of communication, we model this channel by the conditioned probabilities $W_\theta(z,y) = \Prob(Z=z|Y =y), \forall (z,y)\in\setZ_k$. 

\subsubsection{Surjection}
One difficulty of this context is the big size of the set $\setZ_k$ of outcomes under the top-$k$ assumption: $|\setZ_k|= (C)_k$. It is then difficult to establish reliable statistics about the transition matrix $W_\theta$ which is as large as $(C)_k\times(C)_k$.

When working with top-$k$ outputs, Alice resorts to a surjection $\sur_k: \setZ_k\mapsto \set{S}_k$ with $\set{S}_k:=\{0,1,\ldots, k\}$.
This greatly reduces the set of outcomes. We denote $\tilde{z}=\sur_k(z)$ and $\tilde{y}=\sur_k(y)$. 
Function $\sur_k$ is slightly more complex than suggested by this simple notation.
Indeed, for any input $x$, we assume that Alice has a reference class $c(x)\in\setC$.
If this input $x$ is annotated, $c(x)$ is its ground truth class.
Otherwise, Alice computes the top-1 output of all the models she knows, and takes a majority vote to decide on $c(x)$.
For this piece of data, a model gives $\mod(x) = (c_1,\ldots,c_k)$ and the surjection makes:
\begin{equation}
	\sur_k(\mod(x))=
	\begin{cases}
		j &\text{if } \exists j:\,c_j = c(x)\\
		0 &\textit{otherwise}. 
	\end{cases}
	\label{eq:Surjection}
\end{equation}
In words, $\sur_k(\mod(x))$ is the rank of the reference class in the top-$k$ output or $0$ if the reference class is not returned.
In the end, Alice uses a transmission matrix $(W_\theta(\tilde{z},\tilde{y}))$ which is only $(k+1)\times(k+1)$.

\subsection{Detection $(\det,\family,\setA\subsetneq\setB,k)$}
For the detection task, Alice first makes the following hypothesis:
The black-box is a variant of the vanilla model $\mod\in\setA$.
This variant may be the identity ($\bb=\mod$), or a variant she knows, or a variant she does not know.

Contrary to the previous section, Alice randomly chooses $L$ inputs  $(X_1,\ldots,X_L)\subset\setX$ to query the black-box 
and compares the observations $(\tZ_{1},\ldots,\tZ_{L})$ to the outputs she knows $(\tY_{1},\ldots,\tY_{L})$, with
$\tZ_\ell \defi \sur_k(\bb(X_\ell))$, $\tY_\ell \defi \sur_k(\mod(X_\ell)), \forall \ell\in\ent{L}$.
We use capital letters here to outline that these are random variables since Alice randomly chooses the inputs.

There are two difficulties: i) to gauge the distance between the outputs observed from the black-box and from model $\mod$ (see Sect.~\ref{sec:DiscDist})
and ii) to randomly sample informative inputs from the set $\setX$ (see Sect.~\ref{sec:selection}).  

\subsubsection{A discriminative distance}
\label{sec:DiscDist}
Alice is indeed testing two hypothesis:
\begin{itemize}
	\item $\mathcal{H}_1$: The black-box is a variant of model $\mod$. There is a dependence between $\tZ$ and $\tY$ which is captured by the statistical model of the variant:
	$$\Prob_1(\tZ = \tz, \tY =\ty) \defi W_\theta(\tz,\ty)\Prob(\tY=\ty).$$
	\item $\mathcal{H}_0$: The black-box is not a variant of model $\mod$. There is no statistical dependence and
	$$\Prob_0(\tZ = \tz, \tY =\ty) \defi \Prob(\tZ = \tz)\Prob(\tY =\ty).$$
\end{itemize}
The well-celebrated Neyman-Pearson test is the optimal score for deciding which hypothesis holds. For $L$ independent observations, it writes as
\begin{equation}
	\label{eq:NP}
	s = \sum_{j=1}^L \log\frac{\Prob_1(\tZ = \tz_j, \tY =\ty_j)}{\Prob_0(\tZ = \tz_j, \tY =\ty_j)}=\sum_{j=1}^L \log\frac{W_\theta(\tz_j,\ty_j)}{\Prob(\tZ = \tz_j)}.
\end{equation}
We introduce the empirical joint probability distribution defined by
\begin{equation}
	\hat{P}_{\tZ,\tY}(\tz,\ty) \defi L^{-1} |\{j\in\ent{L}: \tz_j=\tz\mbox{ and } \ty_j=\ty\}|
\end{equation}
in order to rewrite~\eqref{eq:NP} as
\begin{equation}
	s = L\sum_{(\tz,\ty)\in\set{S}_k^2} \hat{P}_{\tZ,\tY}(\tz,\ty) \log\frac{W_\theta(\tz,\ty)}{\Prob(\tZ = \tz)}.
\end{equation}
This formalization is not tractable because $W_\theta$ is not known: Alice does not know which variant $\theta$ is in the black-box, and indeed it might be an unknown variant. Yet, it guides us to a more practical score function, the empirical mutual information: 
\begin{equation}
	\hat{I}(\tZ,\tY)\defi \sum_{(\tz,\ty)\in\set{S}_k^2} \hat{P}_{\tZ,\tY}(\tz,\ty) \log\frac{\hat{P}_{\tZ,\tY}(\tz,\ty)}{\hat{P}_{\tZ}(\tz)\hat{P}_{\tY}(\ty)},
\end{equation} 
with the empirical marginal  probabilities:
\begin{equation}
	\hat{P}_{\tZ}(\tz) \defi \sum_{\ty\in\set{S}_k} \hat{P}_{\tZ,\tY}(\tz,\ty),\quad
	\hat{P}_{\tY}(\ty) \defi \sum_{\tz\in\set{S}_k} \hat{P}_{\tZ,\tY}(\tz,\ty).
\end{equation}
In words, the model of the distributions $(\Prob_0,\Prob_1)$ is replaced with empirical frequencies learned on the fly.
Resorting to the empirical mutual information to decode transmitted messages in digital communication is known as Maximum Mutual Information (MMI), recently proven universally optimal~\cite{Tamir:2020va}.

The empirical mutual information is a kind of similarity (the bigger, the more $\tZ$ looks like $\tY$).
Its value lies in the interval $[0, \min(\hat{H}(\tZ),\hat{H}(\tY))]$ with the empirical entropy given by:
\begin{equation}
	\hat{H}(\tZ) \defi -\sum_{\tz} P_{\tZ}(\tz) \log P_{\tZ}(\tz).
\end{equation}
We prefer dealing with a normalized distance and we introduce:
\def \dist {D_L}
\begin{equation}
	\dist(\bb,\mod)\defi 1 - \frac{\hat{I}(\tZ,\tY)}{\min(\hat{H}(\tY),\hat{H}(\tZ))}\in[0,1].
\end{equation}
This defines the distance between the models $\bb$ and $\mod$ respectively producing $\tZ$ and $\tY$.
Indeed, Fig.~\ref{fig:Appetizer} in the introduction is a t-SNE graphical representation extracted from such pairwise distances between models in $\set{B}$. 
For instance, let us consider two extreme scenarios:
\begin{itemize}
	\item The model $\mod$ is in the black-box so that $\tz_j=\ty_j$, $\forall j\in\ent{L}$. Then $P_{\tZ,\tY}(\tz,\ty)=1$ if $\tz=\ty$, and 0 otherwise, producing 
	$\dist(\bb,\mod) = 0.$
	\item The black-box and model $\mod$ yield independent outputs so that $P_{\tZ,\tY}(\tz,\ty) = P_{\tZ}(\tz)P_{\tY}(\ty)$, then
	$\dist(\bb,\mod)= 1.$
\end{itemize}

In the end, Alice deemed the hypothesis $\mathcal{H}_1$ as being true when the distance is small enough:
$\dist(\bb,\mod) < \tau\to\mathcal{H}_1$ is true. 
Alice makes two kinds of errors:
\begin{itemize}
	\item False positive: $\dist(\bb,\mod) < \tau$ whereas $\mathcal{H}_1$ is false.
	\item False negative: $\dist(\bb,\mod)\geq \tau$ whereas $\mathcal{H}_1$ is true.
\end{itemize}
Alice sets the threshold $\tau$ such that the probability of false positive is lower than a required level $\alpha$.
The converse, \ie controlling the probability of false negative, is an illusion.
Appendix~\ref{app:lower} shows for instance that there is no way to theoretically upper bound the distance between a variant and its original model, even if both of them share good accuracy. Our working assumption is that this mutual information is indeed large enough for a reliable hypothesis test and the experimental work confirms this in Sect.~\ref{sec:ExpWork2}.

\subsubsection{Selection of inputs}
\label{sec:selection}
The empirical mutual information is a consistent estimator of the mutual information which depends on the channel transition matrix $W_\theta$ and the input probability distribution $P_{\tY}$.
A result of the theory of communication is that for a given transmission channel, there is an input probability which maximises the mutual information.
This is of utmost importance in order to design a communication system achieving the channel capacity as defined by C.E. Shannon.
In our framework, this would make the distance between a model and its variant closer to 0 likely avoiding a false negative.

However, this idea is not applicable to our scheme because Alice may know a plurality of variants, each of them leading to a different optimal input distribution.
The black-box may also contain an unknown variant excluding any optimization.

Yet, when Alice randomly chooses the inputs, she has the feeling that these inputs must not be too easy to be classified otherwise any model outputs the same prediction. This is not discriminative of the presence of a given model in the black-box and it may lead to a false positive.
On the other hand, these inputs must not be too hard to be classified neither otherwise the prediction tends to be random, destroying the correlation between a model and its variant. This may lead to a false negative.

The experimental work investigates several selection mechanisms of the inputs.
All of them amount to randomly pick inputs from a subset $\setX^\prime$ of $\setX$.
\def\setXa{All}
\def\setXb{50/50}
\def\setXc{30/70}
\def\setXd{Entropy} 
\begin{itemize}
	\item \setXa. There is indeed no selection and  $\setX^\prime=\setX$.
	\item \setXb. Alice's hypothesis concerns a family of variants derived from a vanilla model $m$. $\setX^\prime$ is composed of 50\% of inputs well classified by $m$ (\ie $\mod(x) = c(x)$), 50\% inputs for which $\mod(x)\neq c(x)$.
	\item \setXc. The same definition but with 30\% well classified and 70\% wrongly classified by $\mod$.
	\item \setXd. $\setX^\prime$ is composed of the inputs whose top-1 predictions are highly random.
	For a given input, Alice computes the predictions from all the models in $\setA$ and measures the empirical entropy of these predicted labels.
	She then sorts the inputs of $\setX$ by their entropy, and $\setX^\prime$ contains the head of this ranking.
\end{itemize}
The second and third options are dedicated to the detection task since they only need the vanilla model $\mod$ at the root of Alice's hypothesis.
The last selection mechanism demands a long preprocessing step depending on how big the set of models $\setA$ is.
It is dedicated to the identification task.

\subsection{Identification $(\id,\family,\setA\subsetneq\setB,k)$}
The identification task  is nothing more than an extension of the detection.
Instead of a binary hypothesis, Alice is now facing a multiple hypotheses test with $M+1$ choices:
\begin{itemize}
	\item $\mathcal{H}_i$: The black-box is a variant of vanilla model $\mod_i$, with $1\leq i\leq M$,
	\item $\mathcal{H}_{0}$: The black-box is a variant of an unknown model.
\end{itemize}
The usual way is to compute distance $\dist(\bb,\mod_i)$ per vanilla model $\mod_i\in\setA$, and to decide for model $i^\star = \arg\max_{1\leq i\leq M}\dist(\bb,\mod_i)$, if $\dist(\bb,\mod_{i^\star})$ is lower than a threshold, otherwise Alice chooses hypothesis $\mathcal{H}_{0}$.
If a known model is in the black-box, only three events may occur:
\begin{itemize}
	\item Alice makes a correct identification,
	\item Alice can not make any decision, \ie she deems $\mathcal{H}_{0}$ as true.
	\item Alice makes a wrong identification.
\end{itemize}
Again, by fine-tuning the threshold, Alice controls the probability of the last event.
Note that the probability of success is expected to be smaller than for the previous task.
Identification is more difficult since several hypotheses are competing.

\subsubsection{Compound model}
Information theory helps Alice again thanks to an analogy with the communication over a compound channel.
In this communication problem, a message $\mod_i$ has been emitted and transmitted through a channel $W_\theta$.
The receiver knows a compound channel, \ie a set of channels $\{\theta_j\}_{j=1}^{V}\subset\Theta$.
It knows that the received signal has gone through one of them, but it does not know which one.
There exists an optimal decoder for each channel in the set. The receiver just does not know which one to use.
A theoretically grounded decoder is to decode the received signal with each of the decoders and to aggregate this decoding
with a min operator~\cite{Abbe:2010ta}.

The analogy is the following: the inputs go through all the models $\{\mod_i\}$ known by Alice, and the outputs are like messages.
Bob has chosen one model, \ie one of these messages. Yet, Bob uses a variant which emits noisy outputs observable to Alice.
Now, suppose that Alice knows a set of variants in a given family: $\{\Vf(\mod_i,\theta_j)\}_j\subset\family$.
She uses these variants for computing distances $\dist(\bb,\Vf(\mod_i,\theta_j))$ that she aggregates into one distance w.r.t. the family:
\begin{equation}
\label{eq:compound}
\dist(\bb,\family)\defi \min_j \dist(\bb,\Vf(\mod_i,\theta_j)).
\end{equation}
Intuitively, the black-box might be a very degraded version of a model which is indeed `closer' to a milder variant than to the original model $\mod_i$. 

%
%

\def\pp{0.32}
\begin{figure*}[h]
	\centering
	\begin{subfigure}[b]{\pp\linewidth}
		\centering
		\resizebox{\linewidth}{!}{\input{./images/limited_information/n_images_histogram/random/20.pgf}}
	\end{subfigure}
	\begin{subfigure}[b]{\pp\linewidth}
		\centering
		\resizebox{\linewidth}{!}{\input{./images/limited_information/n_images_histogram/random/100.pgf}}
	\end{subfigure}
	\begin{subfigure}[b]{\pp\linewidth}
		\centering
		\resizebox{\linewidth}{!}{\input{./images/limited_information/n_images_histogram/random/1000.pgf}}
	\end{subfigure}
	\begin{subfigure}[b]{\pp\linewidth}
		\centering
		\resizebox{\linewidth}{!}{\input{./images/limited_information/n_images_histogram/entropy_label/20.pgf}}
		\caption{$L=20$ Images}
	\end{subfigure}
	\begin{subfigure}[b]{\pp\linewidth}
		\centering
		\resizebox{\linewidth}{!}{\input{./images/limited_information/n_images_histogram/entropy_label/100.pgf}}
		\caption{$L=100$ Images}
	\end{subfigure}
	\begin{subfigure}[b]{\pp\linewidth}
		\centering
		\resizebox{\linewidth}{!}{\input{./images/limited_information/n_images_histogram/entropy_label/1000.pgf}}
		\caption{$L=1000$ Images}
	\end{subfigure}
	\caption{\label{fig:unknown_n_images_histogram}Histogram of the distance  $\dist(\mod_1,\mod_2)$ when $(\mod_1,\mod_2)\in\family^2(\mod)$ (orange), $(\mod_1,\mod_2)\in\family^2(\mod,\Psi)$ (green), or $\mod_1$ and $\mod_2$ are variants of different vanilla models (red). Inputs randomly sampled in $\setX$ (\textit{top}) or in $\setX^\prime$ -\setXd\ Sect.~\ref{sec:selection}- (\textit{bottom}).}
\end{figure*}

\subsection{Experimental work}
\label{sec:ExpWork2}
The previous experimental work in Sect.~\ref{sec:ExpWork1} considers three kinds of family concerning the black-box as defined in Sect.~\ref{sec:knowledge_alice}.
When the family is a singleton, because $\family=\family(\mod, \{\theta\})$ or $\family=\family(\mod, \Psi)$ and $|\Psi|=1$, then the distance between the black-box and this unique model is exactly zero. This easy case is now excluded to focus on cases where Alice does not know the variant in the black-box. 

Contrary to Sect.~\ref{sec:ExpWork1}, Alice now resorts to statistical tests.
Any distance between models is a random value since the queries are randomly selected.
Our protocol makes 20 measurements of any considered distance thanks to 20 independent inputs samples.

\subsubsection{Assumptions about the statistical model.}
Section~\ref{sec:unknown_assumptions} makes two assumptions about the statistical dependence between the predictions of models in the same family $\family$ and independence when coming from different families.
Figure~\ref{fig:unknown_n_images_histogram} experimentally verifies these working assumptions. 

The distances between two models $\Vf(\mod,\theta)$ and $\Vf(\mod^\prime,\theta^\prime)$ for two vanilla models $\mod$ and $\mod^\prime$ and any variants $(\theta,\theta^\prime)\in\Theta\times\Theta$ are computed. This sums up to 583,740 combinations.
Figure~\ref{fig:unknown_n_images_histogram} shows the histogram of these distance values over 20 bins in red. 
Perfect statistical independence implies a distance equaling 1.
A low number of queried inputs makes the measured distance spread over a large range.
The selection of the inputs has a major impact. When sampled on $\setX$ (first row), the query may be an `easy' input correctly classified by any model.
This spoils statistical independence. When sampled on $\setX^\prime$ containing inputs hardly correctly classified (second row), the distances are closer to one.     

The figure also shows the histogram of distances between models belonging to the same family spanned by a vanilla model $\mod$, be it $\family(\mod, \Psi)$ (same type of variation) or $\family(\mod)$ (any kind of variation).
As explained in App.~\ref{app:lower}, it is not possible to get a non-trivial upper bound of the distance in this case.
We observe that two models from the same type of variation are usually closer.
It is therefore easier to detect or identify families $\family(\mod, \Psi)$ than $\family(\mod)$.

\subsubsection{Detection $(\det,\family,\setA\subsetneq\setB,k)$} 
\label{sec:detection_expe}
The experiment considers all combinations of hypothesis and model put in the black-box. There are 35 vanilla models and 1046 variants.
This makes 35 families of type $\family(\mod)$ with an average of 30 members per family.
This represents 1081 positive cases and 36,754 negative cases.
There are 377 families of type $\family(\mod, \Psi)$ of which 203 with a size bigger than 1. This makes 907 positive cases and 218,536 negative cases.
The detection performances are gauged by the True Positive Rate (TPR) when threshold $\tau$ is set to get a False Positive Rate (FPR) of $5\%$.

\begin{table}[b]
	\caption{\label{tab:tp_detection_unknown} True Positive Rate for $(\det,\family,\setA\subsetneq\setB,1)$ with 100 random queries selected in $\setX^\prime$ (See Sect.~\ref{sec:selection}). Delegate selected as the closest of $\mod$. False Positive Rate is set to $5\%$.}
	\centering
	\begin{tabular}{c|cccc}
		\toprule
		& \setXa & \setXb & \setXc & \setXd\\
		\midrule
		$\family(\mod)$ & $79.4 \pm 2.1$ & $89.2 \pm 1.3$ & $91.1 \pm 1.5$ & $\bf{95.2 \pm 0.5}$ \\
		$\family(\mod, \Psi)$ & $85.4 \pm 0.9$ & $94.1 \pm 0.7$ & $96.6 \pm 0.5$ & $\bf{99.8 \pm 0.1}$ \\
		\bottomrule
	\end{tabular}
\end{table}

\paragraph{Selection of inputs.} Table~\ref{tab:tp_detection_unknown} shows the TPR obtained when the black-box returns only \textit{top-1} decisions.
As expected, the performances for the families $\family(\mod, \Psi)$ are higher.
The selection \setXd~is clearly the best option.
Its drawback is that it needs statistics about the predictions of many vanilla models.
As far as the detection task is concerned, the other selections are to be preferred because they do not require anything else than the predictions of the suspected vanilla model. In the sequel, the selection \setXc~is used for further experiments on the detection task.

\paragraph{The delegate model}
Alice measures a single distance in between the black-box and a delegate model of the hypothesis' family $\family$. 
Which member of the family is the best delegate? 
Three choices are proposed based on the distance to the vanilla model spanning the family: \emph{Close}, \emph{Median}, and \emph{Far}.
For instance, the \emph{Close} option means that the delegate is the closest member in the family to the vanilla model:
\begin{equation}
	\mod_d = \arg \min_{\mod^\prime\in\family} \dist(\mod, \mod^\prime).
\end{equation}
In the case where $\family=\family(\mod)$, the closest member is $\mod$.
It is not the case when $\family=\family(\mod,\Psi)$, because the vanilla model $\mod$ is not in this family.
Recall that the intersection between two families has to be the empty set, otherwise Alice could not distinguish them.

Table~\ref{tab:delegate_selection} evaluates the three options.
Only the 180 families with more than 3 members are considered here.
For smaller families, the three options would give the same delegate.

The delegate greatly influences the results.
The best choice is to select the delegate as lying at the `center' of the family.
It means the \emph{Close} option for the family $\family(\mod)$, which is indeed the vanilla model $\mod$, or the \emph{Median} option for family $\family(\mod,\Psi)$.

\begin{table*}[bt]
	\caption{\label{tab:delegate_selection} True Positive Rate for $(\det,\family,\setA\subsetneq\setB,k)$ and different delegate options with $L=100$ random queries in \setXc, FPR $=5\%$.}
	\centering
	\begin{tabular}{c|cccccc}
		\toprule
		& \multicolumn{3}{c}{$\family(\mod)$} & \multicolumn{3}{c}{$\family(\mod, \Psi)$}\\
		\cmidrule(rl){2-4} \cmidrule(rl){5-7}
		Delegate & top-$1$ & top-$3$ & top-$5$ & top-$1$ & top-$3$ & top-$5$ \\
		\midrule
		Close & $\bf{91.1 \pm 1.0}$ & $\bf{90.3 \pm 1.2}$ & $\bf{88.2 \pm 0.9}$ & $95.5 \pm 0.7$ & $94.4 \pm 0.6$ & $92.8 \pm 0.7$  \\ 
		Median & $79.5 \pm 0.3$ & $80.2 \pm 0.2$ & $78.3 \pm  0.2$ & $\bf{96.9 \pm 0.5}$ & $\bf{97.8 \pm 0.4}$ & $\bf{96.9 \pm 0.5}$  \\
		Far & $28.4 \pm 2.3$ & $33.8 \pm 2.7$ & $33.7 \pm 3.5$ & $84.8 \pm 1.3$ & $88.4 \pm 1.0$ & $97.3 \pm 0.9$  \\
		\bottomrule
	\end{tabular}	
\end{table*}

\paragraph{Top-$k$ observations} The detection is evaluated for top-\textit{k} outputs in Fig.~\ref{fig:unknown_TP_n_images}.
The best results are surprisingly obtained for $k=1$ in Tab.~\ref{tab:tp_detection_per_images}. Our explanation is the following.
The bigger $k$ the richer the model.  
Yet, the empirical mutual information is calculated from $(k + 1)^2$ estimated probabilities.
For a given number of queries, the fewer estimations the more accurate they are.
The top-1 quickly gets very good results close to 100\%, so that the richer models but less accurately estimated for larger $k$ are not competitive.

To summarize, the TPR reaches 95\%  for 160 queries under top-1, 200 under top-3, and 250 for top-5.

\begin{figure}[bt]
	\centering
	\resizebox{0.9\linewidth}{!}{\input{./images/limited_information/detection/30_70-fuzed.pgf}}
	\caption{\label{fig:unknown_TP_n_images} True Positive Rate for $(\det,\family,\setA\subsetneq\setB,k)$ as a function of the number of queries randomly selected in \setXc, FPR $=5\%$, best delegate options for $\family(\mod)$ (dash) and $\family(\mod, \Psi)$ (plain).}
\end{figure}

\begin{table}[t]
	\caption{\label{tab:tp_detection_per_images} True Positive Rate for $(\det,\family,\setA\subsetneq\setB,k)$ with random queries selected with \textit{30/70}, FPR $=5\%$.}
	\centering
	\begin{tabular}{cc|cccc}
		\toprule
		& & \multicolumn{4}{c}{Number of queries}\\
		\multicolumn{2}{c|}{ }& $L=20$ & $L=50$ & $L=100$ & $L=500$ \\
		\midrule
		\multirow{3}{*}{$\family(\mod)$} & top-1 & $\bf{55.0}$ & $\bf{83.4}$ & $\bf{91.1}$ & $\bf{98.7}$ \\
		& top-3 & $47.9$ & $81.4$ & $90.3$ & $97.9$ \\
		& top-5 & $39.3$ & $78.2$ & $88.2$ & $97.2$\\
		\midrule
		\multirow{3}{*}{$\family(\mod, \Psi)$} & top-1 & $\bf{32.0}$ & $84.7$ & $96.9$ & $99.8$ \\
		& top-3 & $31.2$ & $\bf{91.3}$ & $\bf{97.8}$ & $\bf{100}$ \\
		& top-5 & $29.4$ & $84.7$ & $96.9$ & $\bf{100}$ \\
		\bottomrule
	\end{tabular}
\end{table}

\subsubsection{Identification $(\id,\family,\setA\subsetneq\setB,k)$} All conclusions obtained in the previous section are kept. Alice now has for delegate the vanilla model $\mod$ for $\family(\mod)$ and the \emph{Median} model for $\family(\mod, \Psi)$. Images are sampled with \setXd\ as defined in Sect.~\ref{sec:selection}.

\paragraph{Protocol} We divide the identification task into two parts. First, Alice identifies a family $\family(\mod)$.
As explained earlier, she can make an identification ($\mathcal{H}_i$) or abstain ($\mathcal{H}_0$). 
Depending on what is in the black-box, two correct answers are possible.
In the negative case where $\bb\in\family(\mod^\prime)$ but $\mod^\prime\notin\setA$, the correct answer is to abstain.
The threshold $\tau$ is set s.t. the probability of error in a negative case is set to $5\%$.
In other words, Alice minimizes her decision error when she has to abstain.
For this purpose, $\setA$ is now composed of 30 models and the remaining 5 are used to generate the negative cases.
Alice thus computes the distances between $\bb$ and the 30 vanilla models in $\setA$.
We repeat this 20 times where the tier of 5 excluded models is randomly sampled in $\set{P}$. 
In the positive case where $\bb\in\family(\mod_i)$ and $\mod_i\in\setA$, the correct answer is to decide for hypothesis $\mathcal{H}_i$.

In the second part, Alice identifies the variation, knowing she has made a correct identification of  the global family $\family(\mod)$.
In this case, Alice has to identify the correct variation among 6 families $\{\family(\mod, \Psi_j)\}_{j=1:6}$: randomized smoothing, pruning (filter, all, last), JPEG, posterize (See App.~\ref{app:Models}).
Alice thus computes 6 distances based on their delegates and identifies the family $i^\star=\arg\min_j\dist(\bb,\family(\mod, \Psi_j))$. No thresholding is needed here. 
For each family, 20 variants with random parameters and complying with~\eqref{eq:criterion} are created.
This leads to 700 new models tested in the black-box, different from the 1081 models considered so far.

\begin{figure}[tb]	
	\centering
	\resizebox{0.9\linewidth}{!}{\input{./images/limited_information/identification_pure/decision/stop_pure-mutual_distance-images_sorted_entropy_label-delegate_close-unknown_5-identification.pgf}}
	\caption{\label{fig:unknown_identification_pure} Probability distribution for  $(\id,\family(\mod),\setA\subsetneq\setB,1)$  vs. number $L$ of queries.
		Threshold set to have a maximum 5\% errors in negative cases.}
\end{figure}

\begin{figure}[tb]	
	\centering
	\resizebox{0.9\linewidth}{!}{\input{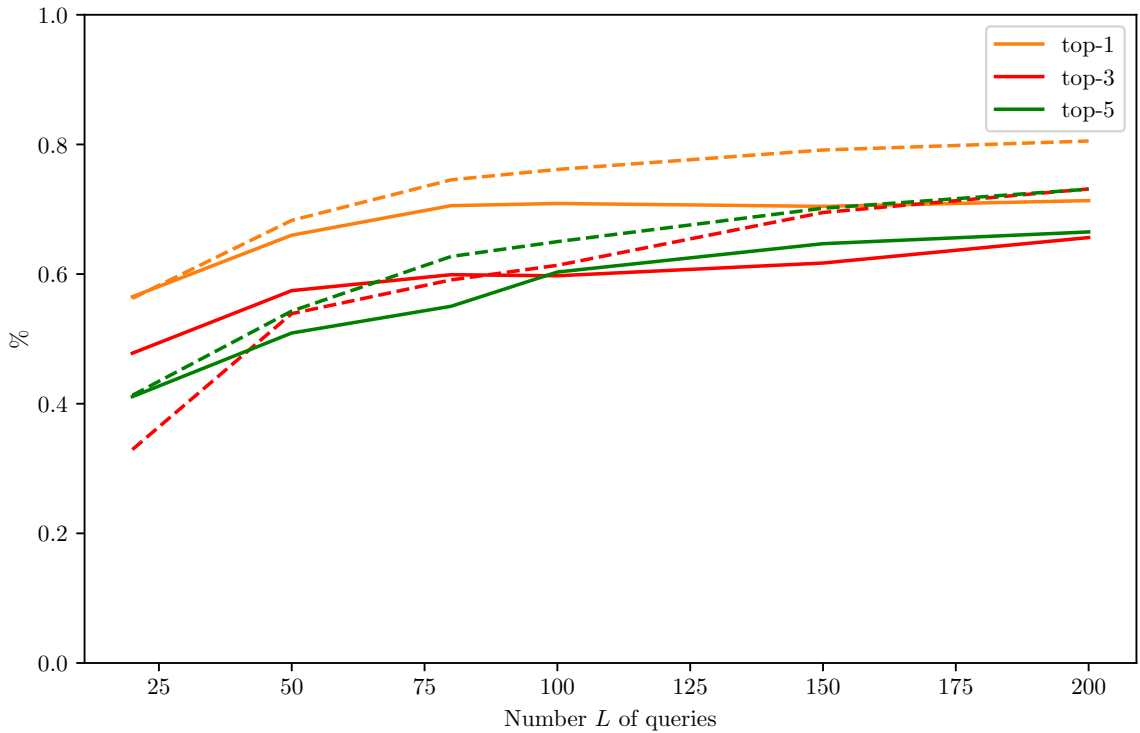}}
	\caption{\label{fig:unknown_identification_variation} Correct Identification Rate for $\family(\mod, \Psi)$ as a function of the number of queries.
	One (plain) or two (dashed) delegates per family.}
\end{figure}

\paragraph{Identifying $\family(\mod)$} 
Alice almost surely identifies the family $\family(\mod)$ of the black-box as shown in Fig.~\ref{fig:unknown_identification_pure} and Tab.~\ref{tab:unknown_identification_topk}.
She reaches her maximum success rate at around 300 queries.
After this amount, no incorrect identification is made but 10\% of abstention remains.
This is due to the thresholding which prevents Alice from misclassification in the negative case.
If no thresholding is done, the success rate reaches 92.8\% within 100 queries and 98.2\% at 500. 

The number of queries is higher than for detection.
For an equivalent level of performance, 4 times more queries are necessary for identification than for detection.
Nevertheless, identification proceeded by sequential detection would take on average 3.000 queries (24 times more w.r.t. detection) as foreseen by~\eqref{eq:IdeExp}.

\paragraph{Identifying $\family(\mod, \Psi)$} With a single delegate, Table~\ref{tab:unknown_identification_topk} and Figure~\ref{fig:unknown_identification_variation} show a rather difficult identification. Variants far from the vanilla model are correctly identified.
The main difficulty comes from the variations that slightly modify the model. These variants are close to $\mod$, which is the center of the cluster $\family(\mod)$ (see Fig.~\ref{fig:tsne_representation_decision_zoom}), therefore it is hard to distinguish them.
The compound~\eqref{eq:compound} with the median and the close delegates yields a boost of 12 points.

\begin{table}[t]
	\caption{\label{tab:unknown_identification_topk} Correct Identification Rate for $(\id,\family,\setA\subsetneq\setB,k)$ with random queries selected with \setXd.} 
	\centering
	\begin{tabular}{cc|ccc}
		\toprule
		& & \multicolumn{3}{c}{Number of queries}\\
		\multicolumn{2}{c|}{ } & $L=50$ & $L=100$ & $L=500$ \\		
		\midrule
		\multirow{3}{*}{\makecell{$\family(\mod)$  \\ delegate = \{close\}}} & top-\textit{1} & \bf{66.7} & \bf{81.3} & \bf{91.8} \\
		& top-\textit{3} & 22.6 & 37.6 & 79.1 \\
		& top-\textit{5} & 21.1 & 48.0 & 82.0 \\
		\midrule
		\multirow{3}{*}{\makecell{$\family(\mod, \Psi)$ \\ delegate = \{median\}}} & top-\textit{1} & 66.0 & 70.9 & 73.7 \\
		& top-\textit{3} & 57.4 & 59.7 & 71.5 \\
		& top-\textit{5} & 50.9 & 60.3 & 69.8 \\
		\midrule
		\multirow{3}{*}{\makecell{$\family(\mod, \Psi)$  \\ delegate = \{close, median\}}} & top-\textit{1} & \bf{68.3} & \bf{76.1} & \bf{82.5} \\
		& top-\textit{3} & 53.9 & 61.4 & 80.6 \\
		& top-\textit{5} & 54.3 & 65.0 & 80.5 \\
		\bottomrule
	\end{tabular}
\end{table}

\paragraph{Top-k observations} The best results are obtained for $k=1$ in Tab.~\ref{tab:unknown_identification_topk} on every task, like for detection. For the family $\family(\mod)$, the information gained by top-$k$ needs too many queries to catch up with the top-1.
For family $\family(\mod, \Psi)$, the difference is smaller. Indeed, top-$k$ with $k\leq3$ gives slightly better results from $\approx$1.000 queries and above.


\def \sensitive{Sensitive Examples~\cite{sensitive}~}
\def \ipguard{IP-Guard~\cite{ipguard}~}

\section{State-of-the-art benchmark}

\subsection{Previous Works}
Since the work of \ipguard, all the fingerprinting papers deal with adversarial examples.
All the papers start with a small collection of benign inputs (except~\cite{wang2021intrinsic} starting from random noise images)
and apply a precise white-box attack like CW. It forges adversarial examples s.t. they lie close to the decision boundaries, which are the signatures of a model.

Two trends are connected to two applications.
The first one deals with the integrity of the model.
In this scenario, Alice makes sure that Bob placed her model in the black-box without any alteration. 
The goal is to sense a \emph{fragile fingerprint} s.t. any modification of the vanilla model is detectable because it changes the fingerprint.
Paper~\cite{sensitive} creates sensitive examples which are adversarial only for the vanilla model.

The second application is \emph{robust fingerprint} as presented so far in this paper.
The followers of \ipguard forge adversarial examples which are more robust in the sense that they remain adversarial for any variation of the model while being more specific to the vanilla model. Paper~\cite{arxiv.2202.08602} proposes to use the universal adversarial perturbations of the vanilla model.
Paper~\cite{lukas2021deep} introduces the concept of conferrable examples, \ie adversarial examples which only transfer to the variations of the targeted model. 
AFA~\cite{ZHAO2020488} activates dropout as a cheap surrogate of variants when forging adversarial examples.
TAFA~\cite{pan2021tafa} extends this idea to other machine learning primitives like regression.

Our take in this paper is that using benign images is sufficient, and
we addressed the fingerprinting problem without the need to rely on
adversarial examples or any other technique to alter images to get
them nearby the frontiers.
Indeed, crafting adversarial examples is rather simple but forging them with extra specificities (fragile or robust to variation) is complex.
It happens that all papers consider small input dimensions like MNIST or CIFAR ($32\times 32$ pixel images); none of them use ImageNet ($224\times224$) except \ipguard.
Also, no paper considers that the inputs can be reformed by a defense (in order to remove an adversarial perturbation before being classified) or detected as adversarial~\cite{Kherchouche:2020wm}.

\subsection{Fragile fingerprinting}
The application considered in~\cite{sensitive} imagines that  Alice wants to detect whether the black-box is exactly $\mod$ and not a variant. This corresponds to our scenario $(\det,\family(\mod,\{\theta\}),\setA=\setB,1)$ where $\theta$ is the identity variation, and $\setA=\family(\mod)$.

We create $L=20$ sensitive examples per model with 200
iterations and two distortion budgets ($\epsilon=8/255$ and $16/255$) using their code\footnote{Sensitive Examples' GitHub: \url{https://github.com/zechenghe/Sensitive_Sample_Fingerprinting}}.
It happens that its performance on ImageNet (reported in Tab.~\ref{tab:benchmark_detection}) is lower than the one reported in~\cite{sensitive} on small input size datasets (like CIFAR). Especially, this scheme can not distinguish the vanilla model and its variants `JPEG' or `Half precision' even with a number of queries ($L=20$) bigger than the one recommended ($L=8$) in~\cite{sensitive}.
Our scheme needs no more than \emph{two} queries and performs perfectly except when pruning the last layer for five out of thirty-five models.

\begin{table*}
	\centering
	\caption{\label{tab:benchmark_detection} False Positive Rate  for the $(\det,\family=\{\mod\},\setA=\setB,1)$ task. }
	\resizebox{\columnwidth}{!}{%
	\begin{tabular}{lc|cccccccccc}
		\toprule
		\multirow{2}{*}{ }&&
		\multicolumn{2}{c}{Finetuning}& Histo. & Random. & \multicolumn{3}{c}{Prune}  & Poster. & Half &  JPEG \\
		\cmidrule(rl){3-4}\cmidrule(rl){7-9}
		& & All & Last & & Smooth. & All & Last & Filter & & Prec. &  \\
		\midrule
		\multirow{2}{*}{Sens. Ex.~\cite{sensitive} $L=20$} & $\epsilon=\frac{8}{255}$  & 18.1 & 20.3 & 0 & 35.7 & 15.4 & 31.5 & 57.4 & 77.0 & 100 & 100 \\
		& $\epsilon=\frac{16}{255}$ & 21.4 & 1.8 & 0 & 31.0 & 10.0 & 21.4 & 48.5 & 63.4 & 97.1 & 97.1 \\
		FBI with $L = 2$ & & \bf{0}  & \bf{0} & \bf{0} & \bf{0} & \bf{0} & \bf{16.9} & \bf{0} & \bf{0} & \bf{0} & \bf{0}\\
		\bottomrule
	\end{tabular}
	}
\end{table*}

\subsection{Robust fingerprinting}
This application is related to our scenario $(\det,\family(\mod),\setA\subsetneq\setB,k)$
\ipguard is the only work showing to be tractable and effective on large input size like in ImageNet.
It leverages several white-box attacks to create adversarial examples. The best results demonstrated in the paper are with the attack CW~\cite{Carlini:2017vr}.  We instead use BP~\cite{bonnet:hal-03467692} because it exhibits similar performances while being much faster (only 50 iterations).
The BP implementation is from GitHub\footnote{Boundary Projection's GitHub: \url{https://github.com/hanwei0912/walking-on-the-edge-fast-low-distortion-adversarial-examples}}.

Table~\ref{tab:benchmark_tp} compares the performances under 100 and 200 queries and top-1 observations.
Any selection of the inputs beats \ipguard. 
Detailed results are reported in App.~\ref{app:benchmark}. 
Some variations are easier to be detected (`precison', `pruning') and the two methods are on par.
On the contrary, randomized smoothing, a variation never considered in the literature, is more difficult.
\ipguard is based on crafting adversarial examples close to the decision boundaries which are greatly `crumpled' by random smoothing.
Not relying on adversarial examples seems to be a clear advantage here.
Our method offers more stability in the results: No variation pulls the TPR below 85\%.

\begin{table}[bt]
	\centering
	\caption{\label{tab:benchmark_tp} Comparison of True Positive Rates for the $(\det,\family(\mod),\setA\subsetneq\setB,1)$ task. FPR set to 5\%.}
	\begin{tabular}{cc|cc}
		\toprule
		&   & \multicolumn{2}{c}{Number of queries} \\
		Fingerprinting scheme & Parameters & $L=100$ & $L=200$\\
		\midrule
		\ipguard & BP \cite{bonnet:hal-03467692} \& 50 iter.   & 66.9 & 72.7 \\
		\\
		\multirow{3}{*}{FBI} & Random   & 79.4 & 91.5\\
		& 30/70   & 91.1 & 97.4 \\
		& Entropy   & \textbf{95.2} & \textbf{97.6} \\
		\bottomrule
	\end{tabular}
\end{table}

\section{Related work}
\label{sec:SOTA}

\paragraph{Model watermarking}
Having reviewed the related work dedicated to the fingerprinting task
in the previous section, we here review another closely related
domain: the \textit{watermarking} of models.

Watermarking is the active counterpart of fingerprinting:
instead of relying on specifics of a fixed model to devise its unique
fingerprint, watermarking modifies the model for which ownership must be proven.
While watermarking is a common practice for decades in the field of image processing~\cite{560421},
it has just recently been incepted into the machine
learning domain.  Uchida et al.~\cite{uchida2017embedding} first
proposed to watermark a deep neural network by embedding it into the weights and biases of the model.
Quickly after this initial proposal, works
instead focused on a black-box model, where the presence of a
watermark can be assessed by Alice from remote interaction with
the suspected deep neural network, just like for fingerprinting.
In~\cite{le2020adversarial}, authors insert
information by altering the decision boundaries through finetuning.
In~\cite{adi2018turning}, authors also retrain the
model to obtain wrong labels for a so-called \textit{trigger set} of
inputs, that constitutes the watermark.
Please refer to~\cite{LI2021171} for a complete overview of the domain. 

Some papers claim that robust fingerprinting could replace watermarking with the clear advantage that no modification of the model is needed~\cite{ZHAO2020488}.
We strongly disagree. No fingerprinting scheme, included ours, brings any formal guarantee on what is in the black-box, just like for human biometric traits.
It is indeed the primary goal of watermarking to guarantee a certain level of trust. 


\section{Conclusion}
The problem of accurate and efficient fingerprinting of valuable modern models is salient.
This paper demonstrates that such a demand can be fulfilled
by just using benign inputs, in not only the classic detection task, but also
in the novel identification task we have introduced.
This implies that we no longer need models in white-box access to compute their fingerprints.

\medskip
We outline the following lessons:\\
In the walled garden setup of Sect.~\ref{sec:known}, less than ten inputs are needed but these are sequentially and carefully selected among a large collection depending on the previous outputs of the black-box. In other words, the key is the interaction between the greedy algorithm and the black-box.
Observing top-1, top-3 or top-5 makes a difference. It is easier to spot inputs that single out a model with richer outputs. 

In the open-world setup of Sect.~\ref{sec:unknown}, hundreds of inputs are necessary but the scheme is not iterative and selection is less crucial.
Surprisingly, observing richer outputs does not yield a gain in this setup.

The identification task is more complex than detection but by a small amount only.
Our identification schemes are much more efficient than the naive sequential search.

The hardest variation of our experimental protocol is randomized smoothing for robust fingerprinting and pruning the last layer for fragile fingerprinting.
It means that the former reduces the statistical dependence of the outputs while the latter hardly perturbs the outputs given by the vanilla model. 

\medskip
One limitation of our work is that it cannot handle classifiers whose
accuracy is almost perfect.  This would happen for too easy
classification setups where the value of models is lower, and
fingerprinting is less critical.  We nevertheless expect future models
and applications to be complex tasks, where reaching good accuracy
levels will remain a struggle.


\newpage
\bibliographystyle{unsrt}
\bibliography{refs}

\begin{thebibliography}{10}

\bibitem{9401119}
Si~Wang and Chip-Hong Chang.
\newblock Fingerprinting deep neural networks - a deepfool approach.
\newblock In {\em 2021 IEEE International Symposium on Circuits and Systems
  (ISCAS)}, pages 1--5, 2021.

\bibitem{ipguard}
Xiaoyu Cao, Jinyuan Jia, and Neil~Zhenqiang Gong.
\newblock Ipguard: Protecting intellectual property of deep neural networks via
  fingerprinting the classification boundary.
\newblock In {\em Proceedings of the 2021 ACM Asia Conference on Computer and
  Communications Security}. Association for Computing Machinery, 2021.

\bibitem{arxiv.2202.08602}
Zirui Peng, Shaofeng Li, Guoxing Chen, Cheng Zhang, Haojin Zhu, and Minhui Xue.
\newblock Fingerprinting deep neural networks globally via universal
  adversarial perturbations, 2022.

\bibitem{pan2021tafa}
Xudong Pan, Mi~Zhang, Yifan Lu, and Min Yang.
\newblock Tafa: A task-agnostic fingerprinting algorithm for neural networks.
\newblock In {\em European Symposium on Research in Computer Security}, pages
  542--562. Springer, 2021.

\bibitem{ZHAO2020488}
Jingjing Zhao, Qingyue Hu, Gaoyang Liu, Xiaoqiang Ma, Fei Chen, and
  Mohammad~Mehedi Hassan.
\newblock Afa: Adversarial fingerprinting authentication for deep neural
  networks.
\newblock {\em Computer Communications}, 150:488--497, 2020.

\bibitem{han2015deep}
Song Han, Huizi Mao, and William~J Dally.
\newblock Deep compression: Compressing deep neural networks with pruning,
  trained quantization and huffman coding.
\newblock {\em arXiv preprint arXiv:1510.00149}, 2015.

\bibitem{ganin2016domain}
Yaroslav Ganin, Evgeniya Ustinova, Hana Ajakan, Pascal Germain, Hugo
  Larochelle, Fran{\c{c}}ois Laviolette, Mario Marchand, and Victor Lempitsky.
\newblock Domain-adversarial training of neural networks.
\newblock {\em The journal of machine learning research}, 17(1):2096--2030,
  2016.

\bibitem{sensitive}
Zecheng He, Tianwei Zhang, and Ruby Lee.
\newblock Sensitive-sample fingerprinting of deep neural networks.
\newblock In {\em 2019 IEEE/CVF Conference on Computer Vision and Pattern
  Recognition (CVPR)}, 2019.

\bibitem{meng2017magnet}
Dongyu Meng and Hao Chen.
\newblock Magnet: a two-pronged defense against adversarial examples.
\newblock In {\em Proceedings of the 2017 ACM SIGSAC conference on computer and
  communications security}, pages 135--147, 2017.

\bibitem{Tamir:2020va}
Ran Tamir and Neri Merhav.
\newblock The mmi decoder is asymptotically optimal for the typical random code
  and for the expurgated code, 2020.

\bibitem{Abbe:2010ta}
Emmanuel Abbe and Lizhong Zheng.
\newblock Linear universal decoding for compound channels.
\newblock {\em IEEE Transactions on Information Theory}, 56(12):5999--6013,
  2010.

\bibitem{wang2021intrinsic}
Siyue Wang, Pu~Zhao, Xiao Wang, Sang Chin, Thomas Wahl, Yunsi Fei, Qi~A Chen,
  and Xue Lin.
\newblock Intrinsic examples: Robust fingerprinting of deep neural networks.
\newblock In {\em British Machine Vision Conference (BMVC)}, 2021.

\bibitem{lukas2021deep}
Nils Lukas, Yuxuan Zhang, and Florian Kerschbaum.
\newblock Deep neural network fingerprinting by conferrable adversarial
  examples.
\newblock In {\em International Conference on Learning Representations}, 2021.

\bibitem{Kherchouche:2020wm}
Anouar Kherchouche, Sid~Ahmed Fezza, Wassim Hamidouche, and Olivier
  D{\'e}forges.
\newblock Detection of adversarial examples in deep neural networks with
  natural scene statistics.
\newblock In {\em 2020 International Joint Conference on Neural Networks
  (IJCNN)}, pages 1--7, 2020.

\bibitem{Carlini:2017vr}
Nicholas Carlini and David Wagner.
\newblock Towards evaluating the robustness of neural networks.
\newblock In {\em 2017 IEEE Symposium on Security and Privacy (SP)}, pages
  39--57, 2017.

\bibitem{bonnet:hal-03467692}
Benoit Bonnet, Teddy Furon, and Patrick Bas.
\newblock {Generating Adversarial Images in Quantized Domains}.
\newblock {\em {IEEE Transactions on Information Forensics and Security}},
  2022.

\bibitem{560421}
M.D. Swanson, Bin Zhu, and A.H. Tewfik.
\newblock Transparent robust image watermarking.
\newblock In {\em Proceedings of 3rd IEEE International Conference on Image
  Processing}, volume~3, pages 211--214 vol.3, 1996.

\bibitem{uchida2017embedding}
Yusuke Uchida, Yuki Nagai, Shigeyuki Sakazawa, and Shin'ichi Satoh.
\newblock Embedding watermarks into deep neural networks.
\newblock In {\em Proceedings of the 2017 ACM on International Conference on
  Multimedia Retrieval}, pages 269--277, 2017.

\bibitem{le2020adversarial}
Erwan Le~Merrer, Patrick Perez, and Gilles Tr{\'e}dan.
\newblock Adversarial frontier stitching for remote neural network
  watermarking.
\newblock {\em Neural Computing and Applications}, 2020.

\bibitem{adi2018turning}
Yossi Adi, Carsten Baum, Moustapha Cisse, Benny Pinkas, and Joseph Keshet.
\newblock Turning your weakness into a strength: Watermarking deep neural
  networks by backdooring.
\newblock In {\em 27th USENIX Security Symposium (USENIX Security 18)}, pages
  1615--1631, 2018.

\bibitem{LI2021171}
Yue Li, Hongxia Wang, and Mauro Barni.
\newblock A survey of deep neural network watermarking techniques.
\newblock {\em Neurocomputing}, 461:171--193, 2021.

\bibitem{rw2019timm}
Ross Wightman.
\newblock Pytorch image models.
\newblock \url{https://github.com/rwightman/pytorch-image-models}, 2019.

\bibitem{torchvision}
S\'{e}bastien Marcel and Yann Rodriguez.
\newblock Torchvision the machine-vision package of torch.
\newblock In {\em Proceedings of the 18th ACM International Conference on
  Multimedia}. Association for Computing Machinery, 2010.

\bibitem{Raff_2019_CVPR}
Edward Raff, Jared Sylvester, Steven Forsyth, and Mark McLean.
\newblock Barrage of random transforms for adversarially robust defense.
\newblock In {\em Proceedings of the IEEE/CVF Conference on Computer Vision and
  Pattern Recognition (CVPR)}, June 2019.

\bibitem{DBLP:journals/corr/abs-1803-10840}
Uri Shaham, James Garritano, Yutaro Yamada, Ethan Weinberger, Alex Cloninger,
  Xiuyuan Cheng, Kelly~P. Stanton, and Yuval Kluger.
\newblock Defending against adversarial images using basis functions
  transformations.
\newblock {\em CoRR}, abs/1803.10840, 2018.

\bibitem{guo2018countering}
Chuan Guo, Mayank Rana, Moustapha Cisse, and Laurens van~der Maaten.
\newblock Countering adversarial images using input transformations.
\newblock In {\em International Conference on Learning Representations}, 2018.

\bibitem{10.1145/3219819.3219910}
Nilaksh Das, Madhuri Shanbhogue, Shang-Tse Chen, Fred Hohman, Siwei Li,
  Li~Chen, Michael~E. Kounavis, and Duen~Horng Chau.
\newblock Shield: Fast, practical defense and vaccination for deep learning
  using jpeg compression.
\newblock In {\em Proceedings of the 24th ACM SIGKDD International Conference
  on Knowledge Discovery; Data Mining}, KDD '18. Association for Computing
  Machinery, 2018.

\bibitem{e22111201}
Anibal Pedraza, Oscar Deniz, and Gloria Bueno.
\newblock Approaching adversarial example classification with chaos theory.
\newblock {\em Entropy}, 22(11), 2020.

\bibitem{feature_squeezing}
Weilin Xu, David Evans, and Yanjun Qi.
\newblock Feature squeezing: Detecting adversarial examples in deep neural
  networks.
\newblock {\em Proceedings 2018 Network and Distributed System Security
  Symposium}, 2018.

\bibitem{DBLP:journals/corr/abs-1902-02918}
Jeremy Cohen, Elan Rosenfeld, and Zico Kolter.
\newblock Certified adversarial robustness via randomized smoothing.
\newblock In {\em ICML}, 2019.

\bibitem{lecuyer2019certified}
Mathias Lecuyer, Vaggelis Atlidakis, Roxana Geambasu, Daniel Hsu, and Suman
  Jana.
\newblock Certified robustness to adversarial examples with differential
  privacy.
\newblock In {\em S\&P}, 2019.

\bibitem{DBLP:journals/corr/abs-1809-03113}
Bai Li, Changyou Chen, Wenlin Wang, and Lawrence Carin.
\newblock Second-order adversarial attack and certifiable robustness.
\newblock 2019.

\bibitem{maho:hal-03591421}
Thibault Maho, Teddy Furon, and Erwan Le~Merrer.
\newblock {Randomized Smoothing under Attack: How Good is it in Pratice?}
\newblock In {\em {ICASSP 2022 - IEEE International Conference on Acoustics,
  Speech and Signal Processing}}. {IEEE}.

\bibitem{https://doi.org/10.48550/arxiv.1608.08710}
Hao Li, Asim Kadav, Igor Durdanovic, Hanan Samet, and Hans~Peter Graf.
\newblock Pruning filters for efficient convnets, 2016.

\bibitem{zhang:hal-02931493}
Hanwei Zhang, Yannis Avrithis, Teddy Furon, and Laurent Amsaleg.
\newblock {Walking on the Edge: Fast, Low-Distortion Adversarial Examples}.
\newblock {\em IEEE TIFS}, 2020.

\end{thebibliography}

\clearpage
\appendix

\section{Notations}
Notations are summarized in Tab.~\ref{tab:notations}.

\begin{table}[h]
\caption{Notations}
\begin{center}
\begin{tabular}{l|l}
\toprule
$\det$ & the task of detection \\
$\id$ & the task of identification \\
$\mod$ & a vanilla model as listed in Tab.~\ref{tab:setA} \\
$\Vf(\mod,\theta)$ & variant obtained by applying procedure $\theta$ on $\mod$ \\
$\Theta$ & set of all variation procedures and parameters \\
$\acc(\mod)$ & accuracy of model $\mod$ \\
$\bb$ & the model in the black-box \\
$\yb{}$ & top-$k$ output of the black-box \\
$\setC$ & the set of classes labeled from 1 to $C$  \\
$\setZ_k$ & the set of top-$k$ outcomes \\
$(C)_k$ &  falling factorial $(C)_k\defi C(C-1)\ldots(C-k+1)$ \\
$\setB$ & set of models Bob can create as defined in~\eqref{eq:setB} \\
$\setA$ & set of models known by Alice \\
$\set{P}$ & set of public vanilla models listed in Tab.~\ref{tab:setA} \\
$\family(\mod)$ & family spanned by vanilla model $\mod$~\eqref{eq:Family1}\\
$\family(\mod,\Psi)$ & family spanned by $\mod$ and variation  $\Psi\subset\Theta$~\eqref{eq:Family2}\\
$\setX$ & a collection of benign inputs \\
$N$ & size of $\setX$ \\
$q_{1:\ell}$ & an ordered list of indices in $\ent{N}$ \\
$\ent{N}$ & integers from 1 to $N$ \\
$\set{D}(x)$ & set of outputs given by models in $\set{D}$ for input $x$ \\
$\set{M}(x,y,\set{D})$ & subset of models of $\set{D}$ giving $y$ for input $x$ \\
$(\setA \backslash\family)^{(\ell)}$ & subset of candidate models at step $\ell$~\eqref{eq:Greedy1}\\
$s^{(\ell+1)}(x)$ & score of input $x$ at step $\ell+1$ when $\setA=\setB$ \\
$L^{\poss}, L^{\negg}$ & nb. of queries for a positive / negative detection \\
$\sur_k$ & surjection from $\setZ_k$ to $\set{S}_k:=\{0,1,\ldots, k\}$ \\
$z,y$ & top-k output of black-box $\bb$ or of model $\mod$ \\
$Z,Y$ & random top-k output when the input is random \\
$\tilde{z}, \tZ, \tilde{y}, \tY$ &  similar outputs after the sujection \\
$W_\theta$ & $(k+1)\times(k+1)$ transition matrix \\
$\hat{I}(\tZ,\tY)$ & Empirical mutual information in bits \\
$\hat{H}_{\tZ}(\tz)$ & Empirical entropy in bits \\
$\dist(\mod_1,\mod_2)$ & Distance between models with $L$ queries \\
$\dist(\bb,\family)$ & Distance of the black-box from family $\family$ \\
\bottomrule
\end{tabular}
\end{center}
\label{tab:notations}
\end{table}%

\clearpage

\section{Models and variations}
\label{app:Models}
\subsection{Description of the set of models}
\label{sec:variation_description}
The core of the set $\setA$ contains 35 `off-the-shelves' vanilla models which were trained with supervision for the task ImageNet over one million annotated $224\times224$ pixel images.
These models come from the Timm \cite{rw2019timm} and Torchvision \cite{torchvision} libraries.
Among them, some are very close. For instance, there are 6 versions of EfficientNet-b0.
These models share the same architecture but they result from different training session.
Yet, they are considered as different models and this proves the efficiency of our method.

To forge a variant, we have selected 8 transformations. These are simple procedures easily applicable by Bob.
They move the decision boundary of the model with a limited drop in accuracy.
As such, all of them have already proven themselves as a defense against adversarial examples~\cite{Raff_2019_CVPR}.

\begin{itemize}
	\item \textbf{Identity}: The variant is an exact copy of the model.	
	\item \textbf{Model Precision}: Deep neural networks usually encode weights and biases on 32 bits floating point precision. The Torch class attribute  \texttt{half} is used to reduce the precision to 16 bits. 
	\item \textbf{JPEG Compression}: Before being classified, the input image goes through a JPEG compression. This has been proposed as a defense: JPEG coarsely quantizes the high frequencies while adversarial perturbations are essentially composed of high frequencies. JPEG compression and decompression act as a reformer~\cite{DBLP:journals/corr/abs-1803-10840, guo2018countering, 10.1145/3219819.3219910}. No training is needed because models are robust to JPEG compression.
	The quality factor ranges from 30 to 90 in step of 10.
	
	\item \textbf{Histogram Equalizer}: This increases the contrast in an image and it has been proposed as a defense against adversarial perturbation in~\cite{Raff_2019_CVPR,e22111201}. We use the function \texttt{transforms.functional.equalize}  from Torchvision.
	
	\item \textbf{Color Depth Reduction}:
	Another defense is to reduce the depth of color channels to less than 8 bits~\cite{feature_squeezing}, here from  3 and 7 bits. The image is posterized with the function \texttt{transforms.functional.posterize} from Torchvision.
	
	\item \textbf{Randomized Smoothing}: Randomized smoothing provides robustness guarantees~\cite{DBLP:journals/corr/abs-1902-02918,lecuyer2019certified,DBLP:journals/corr/abs-1809-03113} and  is also efficient against black-box attacks~\cite{maho:hal-03591421}. We take the Github implementation\footnote{Randomized Smoothing GitHub: https://github.com/locuslab/smoothing} of \cite{DBLP:journals/corr/abs-1902-02918}. The number of samples is set to 100 as recommended in the previous works. The standard deviations $\sigma$ selected are: 0.01, 0.02, 0.04, 0.06, 0.08, 0.1. 
	
	\item \textbf{Finetuning}: Finetuning updates the weights of the model during a new training. We consider finetuning all the layers or only the last one. Finetuning runs over 50 epochs with SGD optimizer.
	
	\item \textbf{Pruning}: Pruning compresses the model by removing the less important weights. It is applied with the function \texttt{nn.utils.prune.l1\_unstructured} from Torch package. It removes the weights with the lowest $\ell_1$ norm. Pruning can be applied on all the layers or just some particular ones. Filter pruning \cite{https://doi.org/10.48550/arxiv.1608.08710} cuts the less important output channels of the convolutional layers. We consider three options with the following fraction of weights removed:
	\begin{itemize}
		\item Pruning All layers: 1\%, 2\%, 3\%, 4\%
		\item Filter Pruning: 10\%, 20\%, 30\%,
		\item Pruning the Last layer: 70\%, 80\%, 90\%, 95\%
	\end{itemize}
\end{itemize}

This makes a total of 33 procedures, most of them being easily and quickly applicable to any vanilla model.
Few of them imply a light retraining which is done with a subset of 50,000 images of the ImageNet validation set.  
We end up with 1189 variants. The accuracy of each of them is measured on the remaining part of the ImageNet validation set.
If the drop of accuracy is bigger than $15\%$ compared to the original model, then the variant is discarded.
Our final collection contains 1081 models and variants.

Alice has a collection of $20.000$ images randomly taken from the ImageNet test set.
These images are not annotated with a ground truth. These were not used for training the models or retraining the variants.  

\begin{table}[h]
\caption{\label{tab:setA} List of the 35 vanilla models and their variants. \textit{all} means that all values listed in App.~\ref{sec:variation_description} comply with~\eqref{eq:criterion}.}
	\resizebox{\columnwidth}{!}{%
	\begin{tabular}{|c|c|c|c|c|c|c|c|c|c|c|c|}
		\hline
		~ & Nb of & \multicolumn{2}{c|}{Finetuning}& Half & Histogram & JPEG & Posterize & \multicolumn{3}{c|}{Prune} & Randomized \\
		& variants & All & Last & Precision & & & & All & Filter & Last & Smoothing \\
		\hline
		$CoaT_{Lite Small}$ & 29 & $\checkmark$ & $\checkmark$ & $\checkmark$ & $\checkmark$ & all & all & 0.01 & all & 0.7, 0.8, 0.9 & all \\ \hline
		$ConViT_{small}$ & 28 & $\checkmark$ & $\checkmark$ & ~ & $\checkmark$ & all & all & 0.01 & all & 0.7, 0.8, 0.9 & all \\ \hline
		$DLA_{102}$ & 30 & $\checkmark$ & $\checkmark$ & $\checkmark$ & ~ & all & 4, 5, 6, 7 & all & all & all & $0.01,0.02, 0.04, 0.06, 0.08$ \\ \hline
		$DLA_{60}$ & 32 & $\checkmark$ & $\checkmark$ & $\checkmark$ & ~ & all & all & all & all & all & all \\ \hline
		$DLA_{68b}$ & 32 & $\checkmark$ & $\checkmark$ & $\checkmark$ & $\checkmark$ & all & all & 0.01, 0.02, 0.03 & all & all & all \\ \hline
		$DPN_{92}$ & 33 & $\checkmark$ & $\checkmark$ & $\checkmark$ & $\checkmark$ & all & all & all & all & all & all \\ \hline
		(Torch) $EfficientNet_{b0}$ & 28 & $\checkmark$ & $\checkmark$ & $\checkmark$ & $\checkmark$ & all & all & 0.01, 0.02 & 0.1, 0.2 & 0.7, 0.8, 0.9 & $0.01,0.02, 0.04, 0.06, 0.08$ \\ \hline
		(TF) $EfficientNet_{b0}$ & 25 & $\checkmark$ & $\checkmark$ & $\checkmark$ & $\checkmark$ & all & all & 0.01 & 0.1, 0.2 & 0.7, 0.8 &  $0.01,0.02, 0.04$\\ \hline
		(TF) $EfficientNet_{b0, AP}$ & 28 & $\checkmark$ & $\checkmark$ & $\checkmark$ & $\checkmark$ & all & all & 0.01, 0.02 & 0.1, 0.2 & 0.7, 0.8 & all \\ \hline
		(TF) $EfficientNet_{b0, NS}$ & 28 & $\checkmark$ & $\checkmark$ & $\checkmark$ & $\checkmark$ & all & all & 0.01, 0.02 & 0.1, 0.2 & 0.7, 0.8, 0.9 & $0.01,0.02,0.04, 0.06,0.08$ \\ \hline
		(TF) $EfficientNetV2_{b0}$ & 30 & $\checkmark$ & $\checkmark$ & $\checkmark$ & $\checkmark$ & all & all & 0.01, 0.02, 0.03 & all & 0.7, 0.8, 0.9 & $0.01,0.02,0.04, 0.06,0.08$ \\ \hline
		$HRNet_{w30}$ & 31 & $\checkmark$ & $\checkmark$ & $\checkmark$ & ~ & all & all & all & all & all &  $0.01,0.02,0.04, 0.06,0.08$\\ \hline
		$LeViT_{128s}$ & 28 & ~ & $\checkmark$ & $\checkmark$ & $\checkmark$ & all & all & 0.01 & 0.1, 0.2 & all & all \\ \hline
		$LeViT_{256}$ & 30 & $\checkmark$ & $\checkmark$ & $\checkmark$ & $\checkmark$ & all & all & 0.01, 0.02 & all & all & all \\ \hline
		$LeViT_{384}$ & 32 & $\checkmark$ & $\checkmark$ & $\checkmark$ & $\checkmark$ & all & all & 0.01, 0.02, 0.03 & all & all & all \\ \hline
		(Torch) $ResNet50$ & 28 & $\checkmark$ & $\checkmark$ & $\checkmark$ & ~ & all & 4, 5, 6, 7 & all & all & 0.7, 0.8, 0.9 &  $0.01,0.02,0.04, 0.06$\\ \hline
		$ResNet50_{madry}$ & 30 & $\checkmark$ & $\checkmark$ & $\checkmark$ & ~ & all & all & all & all & 0.7, 0.8, 0.9 &  $0.01,0.02,0.04, 0.06,0.08$ \\ \hline
		$MixNet_l$ & 29 & $\checkmark$ & $\checkmark$ & $\checkmark$ & $\checkmark$ & all & 4, 5, 6, 7 & 0.01, 0.02, 0.03 & all & 0.7, 0.8, 0.9 & $0.01,0.02,0.04, 0.06,0.08$ \\ \hline
		$MixNet_s$ & 25 & $\checkmark$ & $\checkmark$ & $\checkmark$ & ~ & all & 4, 5, 6, 7 & 0.01, 0.02 & 0.1, 0.2 & 0.7, 0.8, 0.9 & $0.01,0.02,0.04, 0.06$ \\ \hline
		$MixNet_{xl}$ & 32 & $\checkmark$ & $\checkmark$ & $\checkmark$ & $\checkmark$ & all & all & all & all & 0.7, 0.8, 0.9 & all \\ \hline
		$MobileNetV2_{120d}$ & 27 & $\checkmark$ & $\checkmark$ & $\checkmark$ & $\checkmark$ & all & all & 0.01, 0.02, 0.03 & 0.1, 0.2 & 0.7, 0.8 & $0.01,0.02,0.04, 0.06$ \\ \hline
		$MobileNetV2_{140}$ & 25 & $\checkmark$ & $\checkmark$ & $\checkmark$ & $\checkmark$ & all & all & 0.01, 0.02 & 0.1 & 0.7, 0.8 & $0.01,0.02,0.04, 0.06$ \\ \hline
		$MobileNetV3_{rw}$ & 25 & $\checkmark$ & $\checkmark$ & $\checkmark$ & ~ & all & 4, 5, 6, 7 & 0.01, 0.02, 0.03 & 0.1, 0.2 & 0.7, 0.8 & $0.01,0.02,0.04, 0.06$ \\ \hline
		$PiT_{s, distilled}$ & 33 & $\checkmark$ & $\checkmark$ & $\checkmark$ & $\checkmark$ & all & all & all & all & all & all \\ \hline
		$RegNetX_{32}$ & 31 & $\checkmark$ & $\checkmark$ & $\checkmark$ & ~ & all & all & all & all & 0.7, 0.8, 0.9 & all \\ \hline
		$ResNet50$ & 32 & $\checkmark$ & $\checkmark$ & $\checkmark$ & $\checkmark$ & all & all & 0.01, 0.02, 0.03 & all & all & all \\ \hline
		$BiT_{R50\times1}$& 32 & $\checkmark$ & $\checkmark$ & $\checkmark$ & $\checkmark$ & all & all & 0.01, 0.02, 0.03 & all & all & all \\ \hline
		$ReXNet_{130}$ & 30 & $\checkmark$ & $\checkmark$ & $\checkmark$ & $\checkmark$ & all & all & 0.01, 0.02, 0.03 & 0.1, 0.2 & 0.7, 0.8, 0.9 & all \\ \hline
		$ReXNet_{150}$ & 31 & $\checkmark$ & $\checkmark$ & $\checkmark$ & $\checkmark$ & all & all & 0.01, 0.02, 0.03 & 0.1, 0.2 & all & all \\ \hline
		$ReXNet_{200}$ & 31 & $\checkmark$ & $\checkmark$ & $\checkmark$ & $\checkmark$ & all & all & 0.01, 0.02, 0.03 & 0.1, 0.2 & all & all \\ \hline
		$Swin_T$ & 28 & $\checkmark$ & $\checkmark$ & $\checkmark$ & $\checkmark$ & all & all & ~ & all & 0.7, 0.8, 0.9 & all \\ \hline
		$ResNeXt_{101, 32x4d}$ & 33 & $\checkmark$ & $\checkmark$ & $\checkmark$ & $\checkmark$ & all & all & all & all & all & all \\ \hline
		$ResNeXt_{50, 32x4d}$ & 33 & $\checkmark$ & $\checkmark$ & $\checkmark$ & $\checkmark$ & all & all & all & all & all & all \\ \hline
		$Twins-PCPVT_{base}$ & 28 & $\checkmark$ & $\checkmark$ & $\checkmark$ & $\checkmark$ & all & all & ~ & all & 0.7, 0.8, 0.9 & all \\ \hline
		$ViT_{base, patch=16}$ & 29 & $\checkmark$ & $\checkmark$ & $\checkmark$ & $\checkmark$ & all & all & 0.01, 0.02 & all & 0.7, 0.8 & all \\ \hline
	\end{tabular}}
\end{table}

\subsection{Distance between models}
\label{app:DistanceModel}

\begin{figure}[t]
	\begin{center}
		\includegraphics[width=\linewidth]{"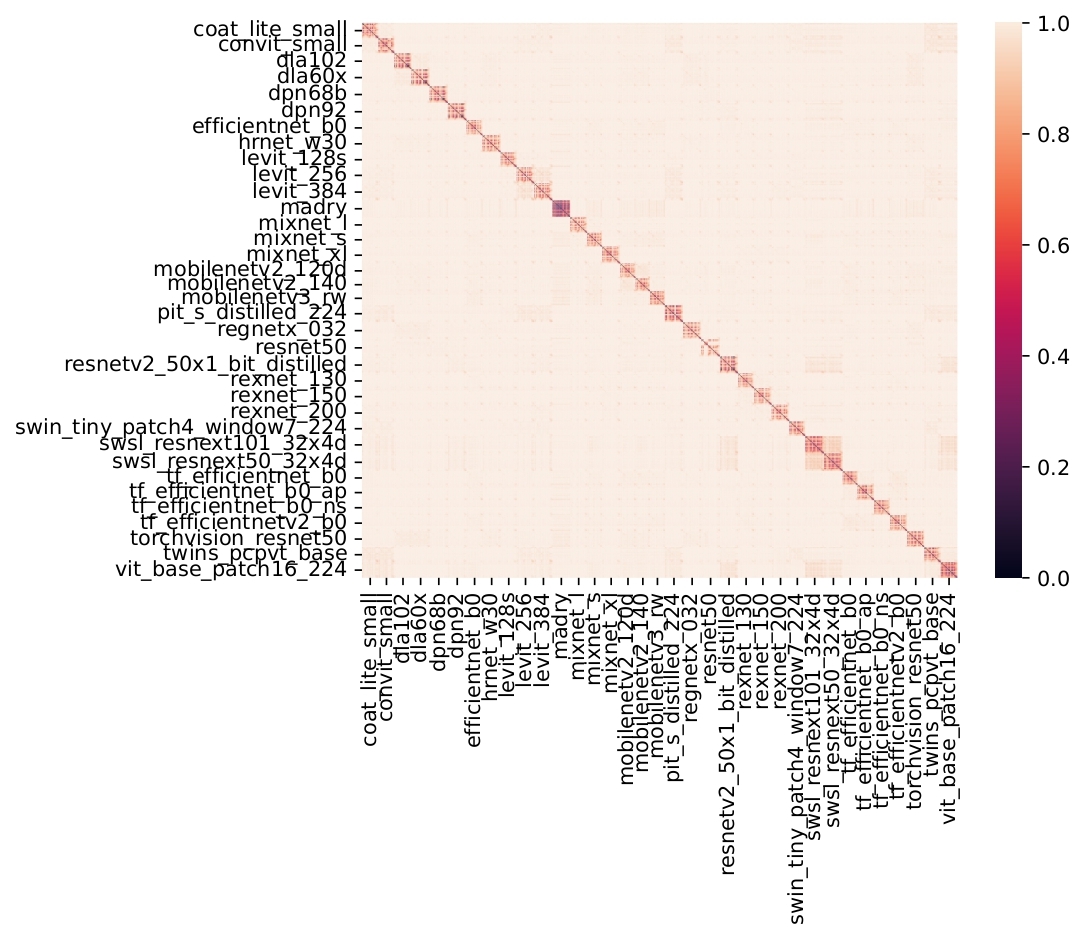"}
	\end{center}
	\caption{\label{fig:mutual_information_distance} Distances between pairs of models.}
\end{figure}

\begin{figure*}
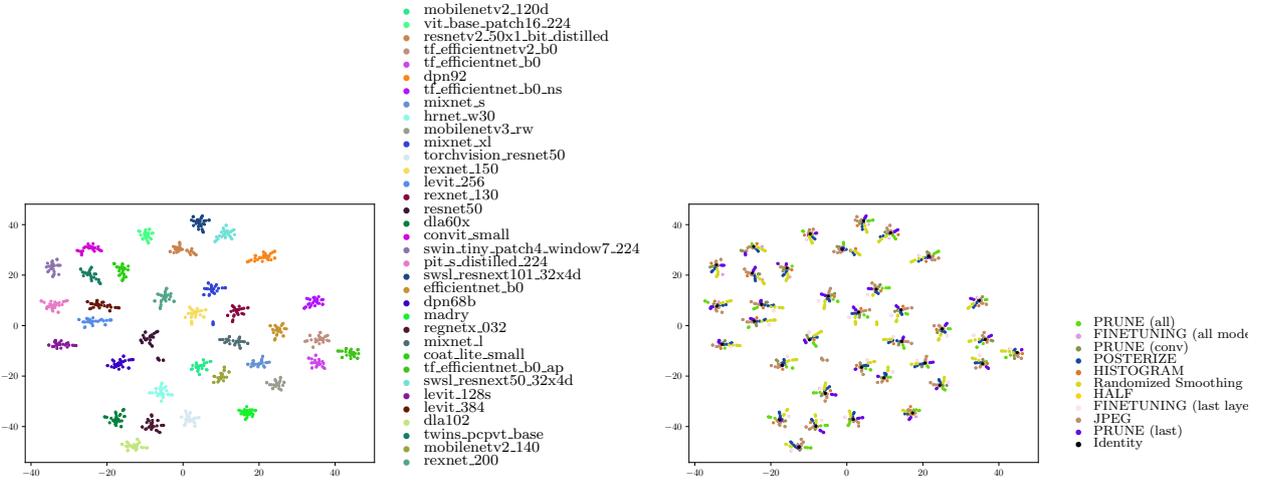

	\begin{center}
		
		\resizebox{0.3\linewidth}{!}{\input{./images/representation/tsne_models_on_decision.pgf}}
		\adjustbox{trim=130 10 110 10, clip, width=0.2\linewidth}{%
			\input{./images/representation/MODELS-legend.pgf}
		}
		\resizebox{0.3\linewidth}{!}{\input{./images/representation/tsne_variants_on_decision.pgf}}
		\adjustbox{trim=130 80 130 100, clip, width=0.15\linewidth}{%
			\input{./images/representation/VARIANTS-legend.pgf}
		}
	
	\end{center}
	\caption{\label{fig:tsne_representation_decision} t-SNE representation on the distances between models. On the left, each color is associated to a vanilla model. On the right, the same representation with a different color code: the color now represents the variation applied to the vanilla models.}
\end{figure*}

The distances between all the pairs of models is shown in Fig.~\ref{fig:mutual_information_distance}.
The block diagonal shows that the distances between variants of the same vanilla models are small. 
This distance matrix is the input of the t-SNE algorithm which creates the 2D representation of Fig.~\ref{fig:tsne_representation_decision}.
We clearly see the cluster of variants centered on each vanilla model.

\begin{figure}[t]
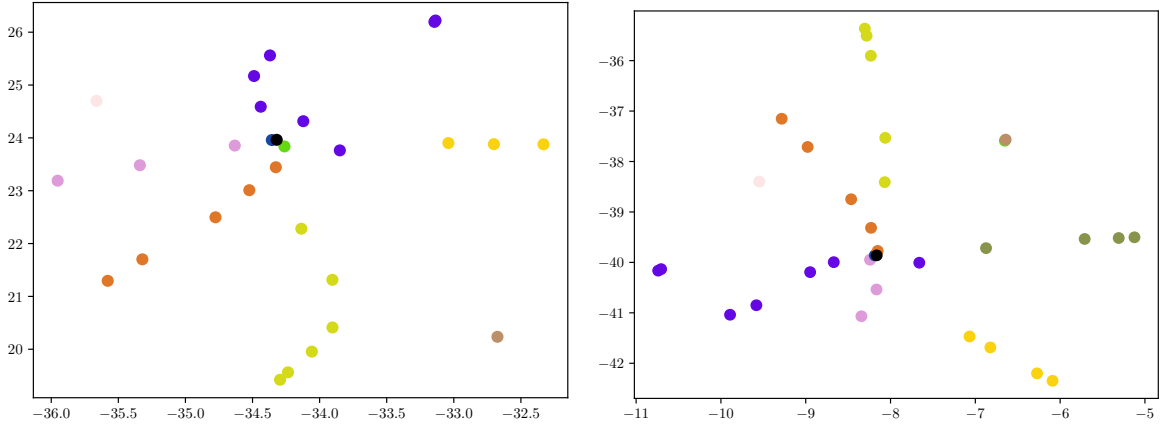

	\begin{center}
		\resizebox{0.45\linewidth}{!}{\input{./images/representation/single_model/swin_tiny_patch4_window7_224-distance_min_matrix-max_drop_-0.15-consensus_0.5-n_images_5000.pgf}}
		\resizebox{0.45\linewidth}{!}{\input{./images/representation/single_model/regnetx_032-distance_min_matrix-max_drop_-0.15-consensus_0.5-n_images_5000.pgf}}
		\caption{t-SNE representation of a single model and his variation. Each color is a variation (see color code of Fig.~\ref{fig:tsne_representation_decision}). The selected vanilla models are $Swin_T$ on the left and  $RegNetX_{32}$.}
		\label{fig:tsne_representation_decision_zoom}
	\end{center}
\end{figure}

\section{Proof of Equation (10)}
\label{sec:ProofEqIdeExp}
Suppose that Bob selects $\bb\in\family_j$.
Alice makes a random permutation $\sigma$ of the $n_{\family}$ families and sequentially tests them.
Then, she spends $L^\negg_i$ queries to discover that the black-box does not belong to $\family_i$, for any $i$ s.t. $\sigma(i)<\sigma(j)$ (\ie the families ranked by the permutation before $\family_j$) and $L^\poss_j$ queries to discover that the black-box is a member of $\family_j$. 
Over all the random permutations, there is statistically one chance out of two that $\family_i$ is ranked before $\family_j$.
On expectation, this makes the following number of queries:
$$
\Exp\left(L^\poss_j\right) + \frac{1}{2}\sum_{i\neq j} \Exp\left(L^\negg_i\right) =
\Exp\left(L^\poss_j\right) + \frac{1}{2}\left(-\Exp\left(L^\negg_j\right) + \sum_{i=1}^{n_\family} \Exp\left(L^\negg_i\right)\right).
$$
We now suppose that Bob select a family uniformly at random, to obtain the following average:
\begin{equation}
\Exp(L) = \frac{1}{n_{\family}}\sum_{j=1}^{n_\family}\Exp\left(L^\poss_j\right)  + \frac{n_\family-1}{2n_\family} \sum_{j=1}^{n_\family}\Exp\left(L^\negg_j\right).
\end{equation}

\section{Lower bound of the distance $\dist$}
\label{app:lower}
This section provides an example of the computation of the mutual information between the outputs of model $\mod$ and the black-box $\bb$.
It considers that these classifiers yield only top-1 outputs.
We assume that the surjection $\sur_1$ is defined in~\eqref{eq:Surjection} w.r.t. the ground truth:
$\tz = 1$ when the black-box succeeds in predicting the ground truth.
The joint probability distribution is denoted by:
$$
\begin{tabular}{c|c|c}
		$\Prob(\tZ,\tY)$ & $\tY=0$ & $\tY=1$ \\
		\hline
		$\tZ=0$ & $a$ & $b$ \\
		$\tZ=1$ & $c$ & $1 - a - b - c$ \\
	\end{tabular}
$$

We suppose that the accuracies of both models are known,
$\acc(\mod) = A$ and $\acc(\bb) = B$, so that:
\begin{eqnarray}
\label{eq:rel1}
\acc(\mod) &=& \Prob(\tY = 1) = 1-a-c = A,\\
\acc(\bb) &=& \Prob(\tZ = 1) = 1-a-b = B.
\label{eq:rel2}
\end{eqnarray}
These equations are constraints reducing the problem with three unknown parameters $(a,b,c)$ to a single one:
From $a$, we can easily deduce $(b,c)$ from the above equations.
Since all these joint probabilities are between 0 and 1, this implies that
\begin{equation}
\max(0,1-(A+B))\leq a\leq \min(1-A,1-B).
\end{equation}

\begin{figure}[h!]
	\centering	
	\resizebox{0.8\linewidth}{!}{\input{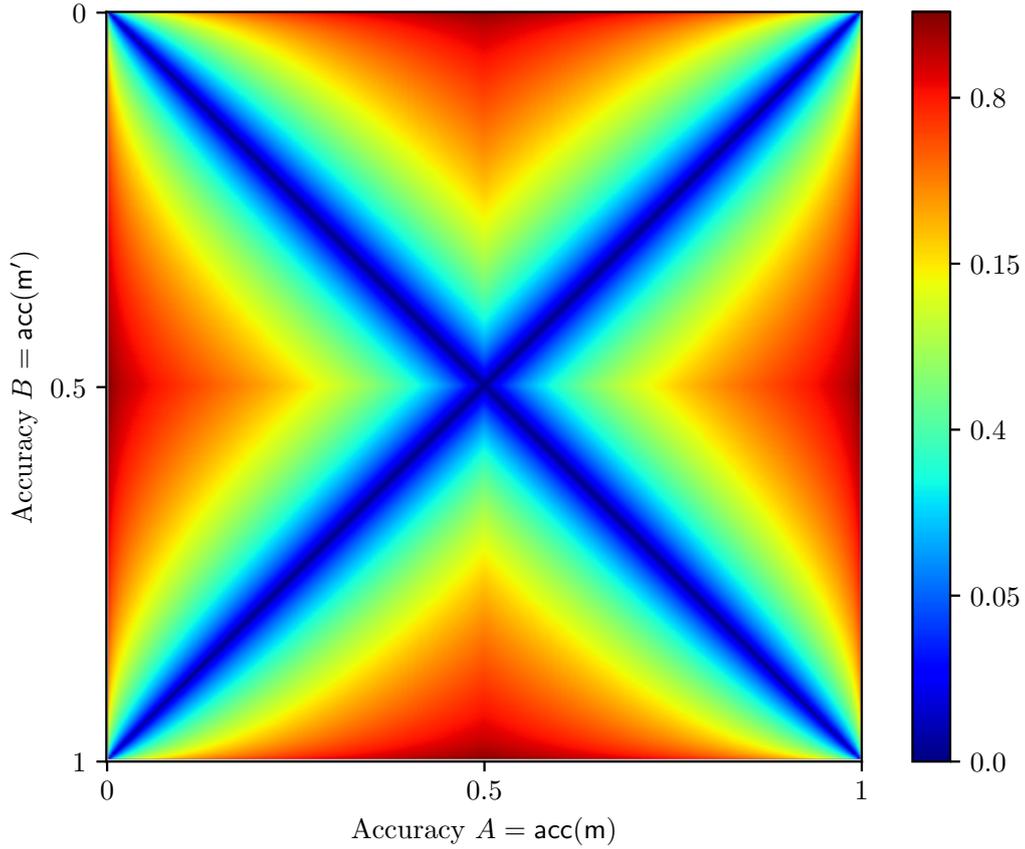}}
	\caption{\label{fig:theory_min_mutual_distance1} Lower bound of the distance between two models of accuracy $A$ and $B$ under top-1.}
\end{figure}

\begin{figure}[h!]
	\centering	
	\resizebox{0.8\linewidth}{!}{\input{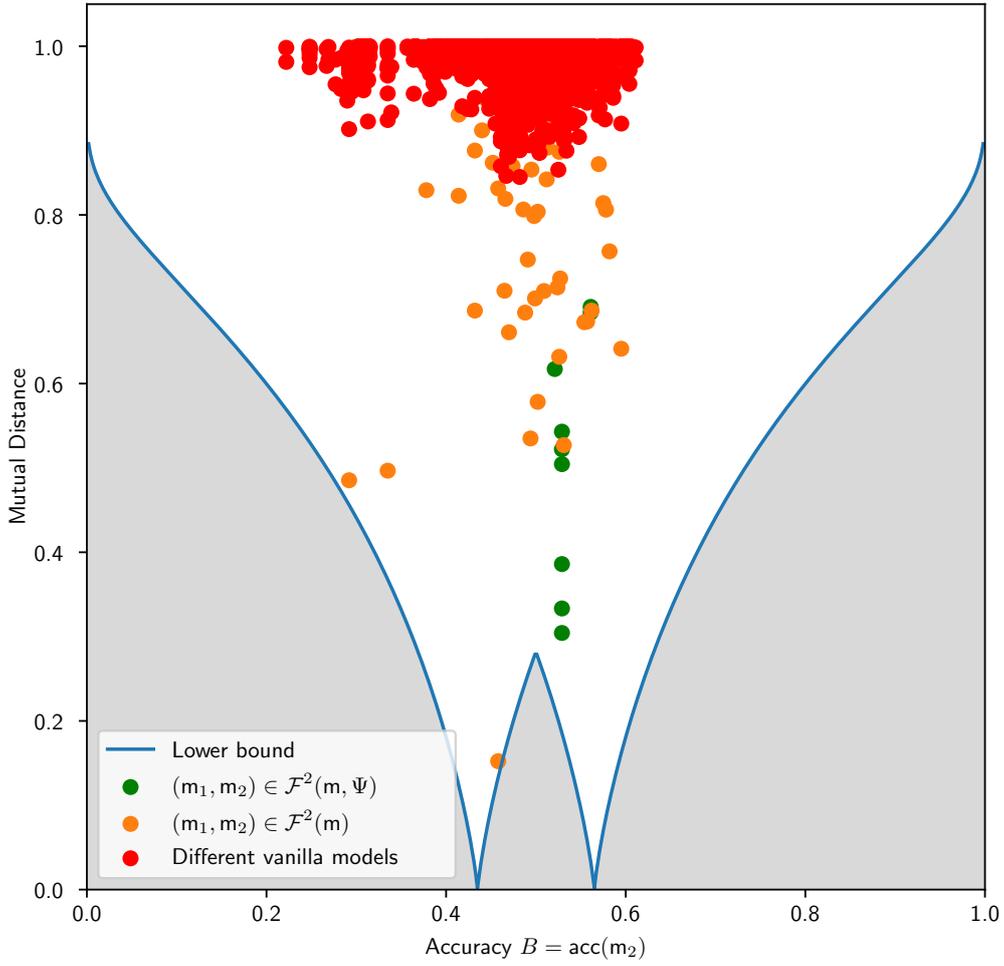}}
	\caption{\label{fig:theory_min_mutual_distance2} Comparison of the theoretical lower bound and the measured distance between models $\mod_1$ and $\mod_2$, when $A=\acc(\mod_1) = 0.45$ and $(\mod_1,\mod_2)\in\family^2(\mod)$ (orange), $(\mod_1,\mod_2)\in\family^2(\mod,\Psi)$ (green), or $\mod_1$ and $\mod_2$ are variants of different vanilla models (red).}
\end{figure}

The mutual information between the outputs $\tZ$ and $\tY$ is given by:
 \begin{eqnarray}
 I(\tZ;\tY) &=& H(\tZ) + H(\tY) - H(\tZ,\tY)\geq 0\\
 &=& h(A) + h(B) - \sum_{(\tz,\ty)\in\{0,1\}^2} f(\Prob(\tZ=\tz,\tY=\ty))\nonumber
 \end{eqnarray}
 where $f(x)\defi  - x\log_2 x$ and $h(p)\defi  f(x)+f(1-x)$ is the binary entropy in bits for $p\in[0,1]$.
Thanks to the constraints~\eqref{eq:rel1} and~\eqref{eq:rel2}, the mutual information is a function of $a$, say $I(\tZ;\tY)=I(a)$.
Its derivative cancels at only one value giving $I(a)=0$ achieved when $\tZ$ and $\tY$ are independent, \ie
$$
a = (1-A)(1-B), \quad b=A(1-B),\quad c=(1-A)B,\quad d=AB.
$$
This is thus a global minimum and the maximum of $I(\tZ;\tY)$ lies on the boundary of the range of $a$.
Suppose that $A\geq B$ and $A+B>1$, then
$$
I(\tZ;\tY) \leq f(A) + \max\left(f(B)-f(A+B-1),f(1-B)-f(A-B)\right)
$$
which is converted into a lower bound of the distance $\dist(\bb,\mod)$
$$
1 - \frac{f(A) + \max\left(f(B)-f(A+B-1),f(1-B)-f(A-B)\right)}{\min(f(A)+f(1-A),f(B)+f(1-B))}.
$$
This function is illustrated in Fig.~\ref{fig:theory_min_mutual_distance1} where we see that $\dist(\bb,\mod)$ may cancel only when $A=B$ or $A=1-B$.
Figure~\ref{fig:theory_min_mutual_distance2} compares this lower bound with the actual measurements for $A=0.45$.

\section{Supplementary results}

\begin{figure*}
	\centering
	\resizebox{0.32\linewidth}{!}{\input{./images/omniscient/detection/score_worst-case-pure.pgf}}
	\resizebox{0.32\linewidth}{!}{\input{./images/omniscient/detection/score_worst-case-variation.pgf}}
	\resizebox{0.32\linewidth}{!}{\input{./images/omniscient/detection/score_worst-case-singleton.pgf}}
	\caption{\label{fig:histogram_all_known_detection_worst_case}Histogram of the number of queries under $(\det,\family,\setA=\setB,k)$  using worst-case score~\eqref{eq:MaxScoreDet} when black-box returns top-1 (\textit{blue}), top-3 (\textit{red}) or top-5 (\textit{green}) decisions. Family considered from left to right: $\family(\mod)$, $\family(\mod,\Psi)$, and $\family(\mod,\{\theta\})$.}
\end{figure*}

\subsection{Known model $\setA=\setB$}
\label{app:known}
Figures~\ref{fig:histogram_all_known_detection_worst_case} and~\ref{fig:histogram_all_known_identification_worst_case}
show detection and identification results when using worst-case score~\eqref{eq:MaxScoreDet}.

\begin{figure*}[h!]
	\centering
	\resizebox{0.32\linewidth}{!}{\input{./images/omniscient/identification/pure-score_worst-case.pgf}}
	\resizebox{0.32\linewidth}{!}{\input{./images/omniscient/identification/variation-score_worst-case.pgf}}
	\resizebox{0.32\linewidth}{!}{\input{./images/omniscient/identification/singleton-score_worst-case.pgf}}
	\caption{\label{fig:histogram_all_known_identification_worst_case}Histogram of the number of queries under $(\id,\family,\setA=\setB,k)$ using worst-case score~\eqref{eq:MaxScoreDet} when black-box returns top-1 (\textit{blue}), top-3 (\textit{red}) or top-5 (\textit{green}) decisions. Family considered from left to right: $\family(\mod)$, $\family(\mod,\Psi)$, and $\family(\mod,\{\theta\})$.}
\end{figure*}

\subsection{Unknown model $\setA\subset\setB$}
\label{app:image_selection}

%
%


Table~\ref{tab:misclassification_proportion} justifies the use of a mixture of well and misclassified inputs under $(\det,\family,\setA\subsetneq\setB,1)$.

\begin{table}[h]
\caption{\label{tab:misclassification_proportion}True Positive Rate for $(\det,\family,\setA\subsetneq\setB,1)$ when queried inputs are a mixture of well or misclassified images.
False Positive Rate is set to 5\%. }
	\centering
	\begin{tabular}{c|ccc}
		\toprule
		Well / Mis-classified & 50 inputs & 100 inputs & 200 inputs \\
		\midrule
		0/100\% & 0.2\% & 0.1\% & 0.0\% \\ 
		5/95\% & \textbf{85.4\%} & 84.8\% & 87.1\% \\ 
		10/90\% & 83.0\% & 84.6\% & 92.8\% \\ 
		15/85\% & 81.2\% & 88.8\% & 94.4\% \\ 
		20/80\% & 82.1\% & 90.3\% & 95.1\% \\ 
		25/75\% & 81.6\% & 91.5\% & 96.6\% \\ 
		30/70\% & 82.7\% & \textbf{91.7\%} & 96.6\% \\ 
		35/65\% & 82.8\% & 91.2\% & \textbf{96.9\%} \\ 
		40/60\% & 80.0\% & 90.8\% & 96.3\% \\ 
		45/55\% & 77.3\% & 90.2\% & 95.6\% \\ 
		50/50\% & 79.7\% & 89.2\% & 95.4\% \\ 
		55/45\% & 75.9\% & 88.8\% & 95.3\% \\ 
		60/40\% & 74.9\% & 87.2\% & 93.1\% \\ 
		65/35\% & 75.2\% & 86.5\% & 92.4\% \\ 
		70/30\% & 71.1\% & 84.7\% & 91.1\% \\ 
		75/25\% & 70.3\% & 85.7\% & 90.9\% \\ 
		80/20\% & 41.8\% & 83.6\% & 89.6\% \\ 
		85/15\% & 17.4\% & 80.9\% & 89.5\% \\ 
		90/10\% & 36.5\% & 30.2\% & 87.5\% \\ 
		95/5\% & 21.7\% & 57.2\% & 78.5\% \\ 
		100/0\% & 0.3\% & 0.6\% & 0.6\% \\ 
		\bottomrule
	\end{tabular}
\end{table}

%
%

\subsection{Benchmark}
\label{app:benchmark}

Table~\ref{tab:benchmark_score_per_variations} details the true positive rates per variation.

\begin{table*}[]
  \caption{\label{tab:benchmark_score_per_variations} True Positive Rate per variation families under $(\det,\family(\mod),\setA\subsetneq\setB,1)$.
	False Positive Rate set to $5\%$ and $L=100$ queries submitted.}
	\centering
	\resizebox{\textwidth}{!}{
	\begin{tabular}{lc|cccccccccc}
		\toprule
		\multirow{2}{*}{Method}& \multirow{2}{*}{Param.} & \multicolumn{2}{c}{Finetune}& \small{Half} & \small{Histo.} & \small{JPEG} & \small{Poster.} & \multicolumn{3}{c}{Prune} & \small{Random.} \\
		\cmidrule(rl){3-4} \cmidrule(rl){9-11}
		& & \small{All} & \small{Last} & \small{Prec.} & & & & \small{All} & \small{Filter} & \small{Last} & \small{Smooth.} \\
		\midrule
		IP-Guard& BP \cite{zhang:hal-02931493} & \multirow{2}{*}{0.5} & \multirow{2}{*}{92.3} & \multirow{2}{*}{\bf{100}} & \multirow{2}{*}{27.3} & \multirow{2}{*}{\bf{100}} & \multirow{2}{*}{9.2} & \multirow{2}{*}{72.7} & \multirow{2}{*}{89.2} & \multirow{2}{*}{\bf{100}} & \multirow{2}{*}{26.1} \\
		\cite{ipguard} & 50 iter. & & & & & & & & & & \\
		& & & & & & & & & & & \\
		\multirow{3}{*}{FBI} & Random & 85.3 & 91.4 & \bf{100} & 77.4 & 92.7 & 88.0 & 69.5 & 91.3 & 90.7 & 48.7 \\
		& 30/70 & 94.1 & \bf{97.1} & \bf{100} & 90.3 & 94.7 & 95.4 & 89.5 & \bf{100} & 95.8 & 79.1 \\
		& Entropy & \bf{94.1} & \bf{97.1} & \bf{100} & \bf{100} & \bf{98.8} & \bf{99.4} & \bf{93.7} & \bf{100}& \bf{100} & \bf{85.6}  \\
		\bottomrule
	\end{tabular}
	}
\end{table*}

\end{document}